# Is Patent Law Technology-Specific?[1]

Dan L. Burk[2] and Mark A. Lemley[3]

> [The software patent cases] stand as a testament to the ability of law to adapt to new and innovative concepts, while remaining true to basic principles.[4]

> Fundamental shifts in technology and in the economic landscape are rapidly making the current system of intellectual property rights unworkable and ineffective. Designed more than 100 years ago to meet the simpler needs of an industrial era, it is an undifferentiated, one-size-fits-all system. Although treating all advances in knowledge in the same way may have worked when most patents were granted for new mechanical devices, today's brainpower industries pose challenges that are far more complex.[5]

Patent law has a general set of legal rules to govern the validity and infringement of patents in a wide variety of technologies. With a very few exceptions, the statute does not distinguish between different technologies in setting and applying legal standards. Rather, those standards are designed to adapt flexibly to new technologies, encompassing "anything under the sun made by man."[6] In theory, then, we have a unified patent system that provides technology-neutral protection to all kinds of technologies.

---

[1]   © 2001 Dan L. Burk & Mark A. Lemley. We thank Kristen Dahling and Colleen Chien for research assistance, and the University of Minnesota Law School and Boalt Hall Law School for research funding.

[2]   Julius E. Davis Professor of Law, University of Minnesota.

[3]   Professor of Law, Boalt Hall, University of California at Berkeley; of counsel, Fish & Richardson P.C.

[4]   AT&T v. Excel Communications, 172 F.3d 1352, 1356 (Fed. Cir. 1999).

[5]   Lester Thurow, *Needed: A New System of Intellectual Property Rights*, **Harv. Bus. Rev.** 95, 95 (Sept.-Oct. 1995).

[6]   Diamond v. Chakrabarty, 447 U.S. 303, 309 (1980) (citing S.Rep.No.1979, 82d Cong., 2d Sess., 5 (1952); H.R.Rep.No.1923, 82d Cong., 2d Sess., 6 (1952)).

Of late, however, we have noticed an increasing divergence between the rules actually applied to different industries. The best examples are biotechnology and computer software. In biotechnology cases, the Federal Circuit has bent over backwards to find biotechnological inventions nonobvious, even if the prior art demonstrates a clear plan for producing the invention. On the other hand, the court has imposed stringent enablement and written description requirements on biotechnology patents that do not show up in other disciplines. In computer software cases, the situation is reversed. The Federal Circuit has essentially excused software inventions from compliance with the enablement and best mode requirements, but in a way that raises serious questions about how stringently it will read the nonobviousness requirements. As a practical matter, it appears, while patent law is technology-neutral in theory, it is technology-specific in application. We provide evidence for this claim in Part I. While our paper focuses on biotechnology and computer software, which present two extreme examples of this phenomenon, our approach has applications in other industries as well, notably chemical inventions and semiconductors.

Part II explains how the application of the same legal standards can lead to such different results in diverse industries. Much of the variance in patent standards is attributable to the use of a legal construct, the "person having ordinary skill in the art" (PHOSITA), to determine obviousness and enablement. The more skill those in the art have, the less information a patentee has to disclose, but the harder it is to find an invention nonobvious. One reading of the biotechnology and computer software cases is that the Federal Circuit believes computer programmers are extremely skilled, while biotechnology experts know very little about their art. We do not challenge the idea that the standards in each industry should vary with the level of skill in that industry. We think the use of the PHOSITA provides needed flexibility for patent law, permitting it to adapt to new technologies without losing its essential character. We fear, however, that the Federal Circuit has not applied that standard properly in either the biotechnology or computer software fields. The court has a perception of both fields that was set in older cases but which does not reflect the modern realities of the industry. The changes in an industry over time present significant structural problems for patent law, both because law is necessarily backward-looking and precedent-bound, and because applying different standards to similar inventions raises concerns about horizontal equity. Nonetheless, we believe the courts must take more care than they currently do to ensure that their assessments of patent validity are rooted in understandings of the technology that were accurate at the time the invention was made.

The question then becomes whether patent law *should* vary from industry to industry. We turn to that question in Part III. The level of skill in the art affects not just patent validity, but also patent scope. Because both claim construction and the doctrine of equivalents turn on the understanding of the PHOSITA in certain circumstances, judgments the court makes about those industries affect the scope of those patents that do issue. In particular, the cases we discuss tend to push the two industries in very different directions: towards a large number of valid biotechnology patents of very limited scope, and a smaller number of valid software patents of quite substantial scope. Given the economics of the two industries, we think it quite possible that this is exactly backwards as a matter of economic policy.

This policy conclusion leads us to question whether patent law should explicitly attempt to tailor protection to the needs of specific industries, as many have suggested. While we draw no definitive conclusions on this score, we do point out a number of risks inherent in such a technology-specific approach. These risks suggest that policy-makers should be cautious about trading our uniform patent system for an industry-specific one. But it may make sense to take economic policy more explicitly into account in designing even general patent rules.

I.     **Heterogeneity in the Patent Law**

A.     **The History of the Uniform Patent System**

A patent statute was one of the first laws Congress passed, in 1790. Since that time, a patent statute has been a constant feature of the U.S. legal landscape.[7] While the nature of the system went through some rather dramatic changes in the first 50 years of the Republic – beginning with a requirement that two cabinet officials must personally review and sign off on any patent,[8] swinging to the other extreme with an automatic registration system subject to caveats[9] -- by 1836 the essential features of the modern patent law were in place.[10] Despite periodic revisions, most recently in 1952, the basic structure of the patent system has remained unchanged for 165 years.

---

[7]  *See, e.g.,* **Bruce Bugbee, The Genesis of American Patent and Copyright Law** 126, 143 (1967); Edward C. Walterscheid, *To Promote the Progress of Useful Arts: American Patent Law and Administration, 1787-1836*, 79 **J. Pat. & Trademark Ofc. Soc'y** 61 (1997) and 80 **J. Pat. & Trademark Ofc. Soc'y** 11 (1998). Even before that time, the U.S. colonies granted patent rights. *See* **Robert P. Merges et al., Intellectual Property in the New Technological Age** 127 (2d ed. 2000).

[8]  This was a feature of the short-lived Patent Act of 1790. *See* Walterscheid, *supra* note __; Edward C. Walterscheid, *Charting a Novel Course: The Creation of the Patent Act of 1790*, 25 **AIPLA Q.J.** 445 (1997).

[9]  The 1793 Act replaced the cumbersome cabinet-level review with a registration system. Under this system, patents were granted without examination unless a competitor or other interested party filed a "caveat" – essentially a request to be notified and given a chance to object if someone patented in a particular field. *See* Walterscheid, *supra* note __, at __.

Technology, of course, has changed dramatically during that time. The "useful arts" envisioned by the Framers were mechanical inventions useful in a primarily agrarian economy. Since that time, the country has gone through a number of periods of dramatic innovation in a wide variety of fields. As late as 1950, though, most inventions were still mechanical in nature. It is only in the last half-century – and to a large extent in the last 25 years, as Allison and Lemley show[11] -- that patent law has lost its primarily mechanical character, branching out into biotechnology, semiconductors, computer hardware and software, electronics, and telecommunications.

What is notable about this history is that the fundamental rules of patent law were set in a world in which inventions were mechanical. Because inventions in the past were far more homogenous than they are today,[12] it made sense to have a unified set of rules for dealing with those inventions. The application of those old rules to new technologies has not been free from controversy. Some have suggested that the unified rules suitable for the old, homogeneous world are no longer appropriate in today's increasingly complex innovative landscape.[13] But without changing the rules themselves, in the last dozen years the Federal Circuit has applied those rules in a way that effectively creates different standards for different industries.[14] In the sections that follow, we examine the treatment of two such industries in detail: computer software and biotechnology.

---

[10] *See* **Merges et al.,** *supra* note __, at 128.

[11] *See* John R. Allison & Mark A. Lemley, *The Growing Complexity of the U.S. Patent System, 1976-1998* (working paper 2001).

[12] *Id*.

[13] *See infra* notes __-__ and accompanying text.

### B. Software Patent Law

Software is patentable today, though it was not always so.[15] The Federal Circuit has moved towards declaring software patentable by fits and starts for years. Finally, with the late-1990s decisions in *State Street Bank*[16] and *AT&T v. Excel*,[17] the court unreservedly admitted software to the pantheon of patentable subject matter. In doing so, the court emphasized that it was deciding only the question of whether software could be patented under section 101.[18] It left the remaining patent validity issues – notably novelty,[19] nonobviousness,[20] and compliance with the disclosure requirements[21] – to be worked out by the courts on a case-by-case basis.[22]

---

[14] Hodges observes that computers and biotechnology are treated differently in the written description cases, though he limits his focus primarily to biotechnology. Robert A Hodges, *Black Box Biotech Inventions: When a Mere "Wish or Plan" Should be Considered an Adequate Description of the Invention*, 17 **Ga. St. U. L. Rev.** 831, 832 (2001). Others have complained that even within industries the standard may not be applied consistently. *See, e.g.,* Glynn S. Lunney Jr., *E-Obviousness*, 7 **Mich. Telecom. & Tech. L. Rev.** 363, 365 & n.13 (2001).

[15] The curious history of the patentability of software is discussed in detail elsewhere. *See, e.g.,* Julie E. Cohen & Mark A. Lemley, *Patent Scope and Innovation in the Software Industry*, 89 **Calif. L. Rev.** 1 (2001); Pamela Samuelson, *Benson Revisited: The Case Against Patent Protection for Algorithms and Other Computer Program-Related Inventions*, 39 **Emory L.J.** 1025, 1033 n.24 (1990); Gregory A. Stobbs, Software Patents (1995); David S. Benyacar, *Mathematical Algorithm Patentability: Understanding the Confusion*, 19 Rutgers Computer & Tech. L.J. 129 (1993); Donald S. Chisum, *The Patentability of Algorithms*, 47 U. Pitt. L. Rev. 959 (1986); Irah H. Donner & J. Randall Beckers, *Throwing Out Baby Benson with the Bath Water: Proposing a New Test for Determining Statutory Subject Matter*, 33 Jurimetrics J. 247 (1993); Lee Hollaar, *Justice Douglas Was Right: The Need For Congressional Action On Software Patents*, 24 AIPLA Q.J. 283 (1996); Allen Newell, *The Models Are Broken, The Models Are Broken!*, 47 U. Pitt. L. Rev. 1023 (1986); Richard H. Stern, *Tales from the Algorithm War: Benson to Iwahashi, It's Déjà Vu All Over Again*, 18 AIPLA Q.J. 371 (1991); Jur Strobos, *Stalking the Elusive Patentable Software: Are There Still Diehr or Was It Just a Flook?*, 6 Harv. J.L. & Tech. 363 (1993); John Swinson, *Copyright or Patent or Both: An Algorithmic Approach to Computer Software Protection*, 5 Harv. J.L. & Tech. 145 (1991); Jonathan N. Geld, Note, *General Does Not Mean Generic -- Shedding Light on In re Alappat*, 4 Tex. Intell. Prop. L.J. 71 (1995); Maximilian R. Peterson, Note, *Now You See It, Now You Don't: Was It a Patentable Machine or an Unpatentable "Algorithm"? On Principle and Expediency in Current Patent Law Doctrines Relating to Computer-Related Inventions,* 64 Geo. Wash. L. Rev. 90 (1995).

[16] State Street Bank & Trust v. Signature Fin. Group, 149 F.3d 1368 (Fed. Cir. 1998).

[17] AT&T Corp. v. Excel Communications, 172 F.3d 1352 (Fed. Cir. 1999).

Section 112 of the Patent Act requires that patentees publish to the world a description of their invention sufficient to enable one of ordinary skill in the art to make and use it, as well as their "best mode" of implementing the invention.[23] Indeed, this disclosure "bargain" between patentees and the public is central to patent policy.[24] Disclosure servers two purposes. First, it permits competitors to make use of the patented invention once the patent expires, ensuring that the invention will ultimately enter the public domain.[25] Second, it enables others to improve on the patented technology during the term of the patent itself, either by "designing around" the patent to produce a non-infringing variant or by developing a better version that, while infringing, is itself entitled to protection.[26]

---

[18] 35 U.S.C. §101.

[19] 35 U.S.C. §102.

[20] 35 U.S.C. §103.

[21] 35 U.S.C. §112 ¶1.

[22] *See State Street*, 149 F.3d at 1375. Indeed, on remand in that case the district court held the patent invalid under section 102. AT&T Corp. v. Excel Communications, 52 U.S.P.Q.2d 1865 (D. Del. 1999).

[23] 35 U.S.C. § 112 ¶1 (1994).

[24] One classic justification for having a patent system is to encourage inventors to disclose their ideas to the public, who will benefit from this new knowledge once the patent expires. Kewanee Oil Corp. v. Bicron Corp., 416 U.S. 470, 489 (1974) (referring to the "federal interest in disclosure" embodied in the patent laws); *see also* Edith Tilton Penrose, The Economics of the International Patent System 31-34 (1951).

[25] Without the disclosure obligation, patentees could conceivably keep the workings of their inventions secret, relying on that secrecy to extend protection even after the patent has expired. *Cf.* Pitney-Bowes v. Mestre, 701 F.2d 1365, 1372 n.12 (11th Cir. 1983) (discussing the policy concerns here).

[26] For a detailed discussion of how the law allocates rights between initial inventors and improvers, see, *e.g.,* Mark A. Lemley, *The Economics of Improvement in Intellectual Property Law*, 75 **Tex. L. Rev.** 989

For software patents, however, a series of recent Federal Circuit decisions has all but eliminated the enablement and best mode requirements. In recent years, the Federal Circuit has held that software patentees need not disclose source or object code, flowcharts, or detailed descriptions of the patented program. Rather, the court has found high-level functional description sufficient to satisfy both the enablement and best mode doctrines.[27] For example, in *Northern Telecom, Inc. v. Datapoint Corp.*,[28] the patent claimed an improved method of entering, verifying, and storing (or batching") data with a special data entry terminal. The district court invalidated certain claims of the patent on the grounds that they were inadequately disclosed under section 112. The Federal Circuit reversed. It held that when claims pertain to a computer program that implements a claimed device or method, the enablement requirement varies according to the nature of the claimed invention as well as the role and complexity of the computer program needed to implement it. Under the facts in this case, the core of the claimed invention was the combination of components or steps, rather than the details of the program the applicant actually used. The court noted expert testimony that various programs could be used to

---

(1997); Robert P. Merges, *Intellectual Property Rights and Bargaining Breakdown: The Case of Blocking Patents*, 62 **Tenn. L. Rev.** 75 (1994); Robert P. Merges & Richard R. Nelson, *On the Complex Economics of Patent Scope*, 90 **Colum. L. Rev.** 839 (1990); Suzanne Scotchmer, *Standing on the Shoulders of Giants: Cumulative Research and the Patent Law*, 4 **J. Econ. Persp.** 29 (1991) Jerry R. Green & Suzanne Scotchmer, *On the Division of Profit in Sequential Innovation*, 26 **RAND J. Econ.** 20 (1995); Suzanne Scotchmer, *Protecting Early Innovators: Should Second-Generation Products be Patentable?*, 27 **RAND J. Econ.** 322 (1996); Howard F. Chang, *Patent Scope, Antitrust Policy, and Cumulative Innovation*, 26 **RAND J. Econ.** 34 (1995); James B. Kobak Jr., *Intellectual Property, Competition Law and Hidden Choices Between Original and Sequential Innovation*, 3 **Va. J. L. & Tech.** 6 (1998); Clarisa Long, *Proprietary Rights and Why Initial Allocations Matter*, 49 **Emory L.J.** 823 (2000).

[27] *See* Fonar Corp. v. General Electric Co., 107 F.3d 1543, 1549 (Fed. Cir. 1997); *see also* Lawrence D. Graham & Richard O. Zerbe, Jr., *Economically Efficient Treatment of Computer Software: Reverse Engineering, Protection, and Disclosure*, 22 Rutgers Computer & Tech. L.J. 61, 96-97 (1996); Anthony J. Mahajan, Note, *Intellectual Property, Contracts, and Reverse Engineering After ProCD: A Proposed Compromise for Computer Software*, 67 Fordham L. Rev. 3297, 3317 (1999).

[28] 908 F.2d 931 (Fed. Cir.), cert. denied, 111 S. Ct. 296 (1990).

implement the invention, and that it would be "relatively straightforward [in light of the specification] for a skilled computer programmer to design a program to carry out the claimed invention."[29] The court continued:

> The computer language is not a conjuration of some black art, it is simply a highly structured language * * * * [T]he conversion of a complete thought (as expressed in English and mathematics, i.e. the known input, the desired output, the mathematical expressions needed and the methods of using those expressions) into a language a machine understands is necessarily a mere clerical function to a skilled programmer.[30]

And in *Fonar v. General Electric*,[31] involving the best mode requirement, the Court explained:

> As a general rule, where software constitutes part of a best mode of carrying out an invention, description of such a best mode is satisfied by a disclosure of the functions of the software. This is because, normally, writing code for such software is within the skill of the art, not requiring undue experimentation, once its functions have been disclosed. It is well established that what is within the skill of the art need not be disclosed to satisfy the best mode requirement as long as that mode is described. Stating the functions of the best mode software satisfies that description test. We have so held previously and we so hold today. Thus, flow charts or source code listings are not a requirement for adequately disclosing the functions of software.[32]

Indeed, the Federal Circuit has gone so far as to hold that patentees can satisfy the written description and best mode requirements for inventions implemented in software even though they do not use the terms "computer" or "software" anywhere in the specification![33] To be sure, in these latter cases it would probably be obvious to one

---

[29] *Id*. at 941-42.

[30] *Id*.

[31] 107 F.3d 1543 (Fed. Cir. 1997).

[32] *Id*. at 1549 (citations omitted).

[33] Robotic Vision Sys., Inc. v. View Eng'g, Inc., 112 F.3d 1163 (Fed. Cir. 1997) (best mode); *In re* Dossel, 115 F.3d 942 (Fed. Cir. 1997) (written description).
    In *White Consolidated Industries, Inc. v. Vega Servo-Control, Inc.*, 713 F.2d 788 (Fed. Cir. 1983), by contrast, the Federal Circuit had invalidated a patent for a machine tool control system which was run by a computer program. Part of the invention was a programming language translator designed to convert an

skilled in the art that the particular feature in question should be implemented in software. Still, it is remarkable that the Federal Circuit is willing to find the enablement requirement satisfied by a patent specification that provides *no* guidance whatsoever on how the software should be written.[34]

It is simply unrealistic to think that one of ordinary skill in the programming field can necessarily reconstruct a computer program given no more than the purpose the program is to perform. Programming is a highly technical and difficult art. But the Federal Circuit's peculiar direction in the software enablement cases has effectively nullified the disclosure obligation in those cases.[35] And since source code is normally kept secret, software patentees generally do not disclose much if any detail about their

---

input program into machine language, which the system could then execute. The patent specification identified an example of a translator program, the so-called "SPLIT" program, which was a trade secret of the plaintiff. The court held that the program translator was an integral part of the invention, and that mere identification of it was not sufficient to discharge the applicant's duty under section 112. The court seemed concerned that maintaining the translator program as a trade secret would allow White to extend the patent beyond the 17-year term then specified in the patent code.

While *White* suggests that it is not sufficient merely to identify the program or its functions, more recent Federal Circuit authority is overwhelmingly to the contrary. *See, e.g., Dossel*, *supra*, at 946 ("While the written description does not disclose exactly what mathematical algorithm will be used to compute the end result, it does state that "known algorithms" can be used to solve standard equations which are known in the art."; finding this sufficient to describe the invention).

[34] One recent decision even found that a specification that provided inconsistent guidance as to how the invention worked was not indefinite. *See* S3 Corp. v. Nvidia Corp., __ F.3d __ (Fed. Cir. Aug. 3, 2001); *compare id*. at __ (__, J., dissenting).

[35] A recent development in Federal Circuit jurisprudence may suggest another source for a robust disclosure obligation, however. The court has recently reinvigorated the written description requirement in section 112 ¶1, not only in biotechnology cases, *e.g.*, Regents of the University of California v. Eli Lilly & Co., 119 F.3d 1559 (Fed. Cir. 1997), but also in cases about mechanical inventions. *E.g.*, Gentry Gallery, Inc. v. Berkline Corp., 134 F.3d 1473 (Fed. Cir. 1998); Hyatt v. Boone, 146 F.3d 1348 (Fed. Cir. 1998). Under those cases, a patent claim may be invalid in certain circumstances if the specification does not expressly describe what the claim covers, even if the specification gave sufficient information to enable the claim. *See also* Johnson & Johnson Assoc. v. R.E. Service Co., 238 F.3d 1347 (Fed. Cir. 2001) (taking en banc sua sponte a case presenting the issue of dedication to the public domain); Maxwell v. J. Baker, 86 F.3d 1098 (Fed. Cir. 1996). We argue below that a broad reading of the written description requirement is largely unique to biotechnology cases. If we are wrong, however, and cases like *Lilly* represent a general rule, it could mean that most software patents will be held invalid for failure to describe the invention in any detail. *But see* In re Dossel, 115 F.3d 942, 946 (Fed. Cir. 1997) (rejecting written description argument in a software case, albeit before the Federal Circuit's more recent cases on the issue).

programs in the patent. Software patentees during the 1980s and early 1990s tended to write their patents in means-plus-function format[36] in order to satisfy the changing dictates of the Federal Circuit's patentable subject matter rules.[37] Lawyers writing patents in such a format have an incentive to describe their invention in the specification in as general terms as possible, since means-plus-function claim elements will be limited to the actual structure disclosed in the specification and equivalents thereof.[38] As a result, there is no easy way to figure out what a software patent owner has built except to reverse engineer the program.[39]

---

[36] *See* 35 U.S.C. §112, ¶6.

[37] For an example, see In re Alappat, 33 F.3d 1526 (Fed. Cir. 1994) (en banc).

[38] *Id*.

[39] On the perils of reverse engineering patented software, see Cohen & Lemley, *supra* note __, at 17-21.
     For discussions of how to satisfy the disclosure requirement in software patents, see Wesley L. Austin, *Software Patents*, 7 **Tex. Intell. Prop. L.J.** 225, 277 (1999); David Bender & Anthony R. Barkume, *Disclosure Requirements for Software-related Patents*, 8 **Computer Law.** 1 (1991); Michael Bondi, *Upholding the Disclosure Requirements of 35 U.S.C. § 112 Through The Submission of Flow Charts with Computer Software Patent Applications*, 5 **Software L.J.** 635 (1992); D.C. Toedt III, *Patents for Inventions Utilizing Computer Software: Some Practical Pointers*, 9 **Computer Law.** 12 (1992) (suggesting disclosure of "pseudocode," i.e., generalized code not written in a particular programming language, to satisfy section 112; and discussing pros and cons of disclosing actual source code). For a policy argument in favor of greater disclosure, see Thomas P. Burke, Note, *Software Patent Protection: Debugging the Current System*, 69 **Notre Dame L. Rev.** 1115, 1158-1160 (1994):
> A software patent without source code is like a law review piece filled with case names but missing citations to case reporters. A person of ordinary skill in legal research might be able to track down the full-text of all the opinions. Marbury v. Madison would be found quicker than a state trial court opinion. But, would anyone think that such a practice was enabling or the best mode? As it is now, the disclosure requirements can be met using such devices as specifications, flowcharts, and pseudo-code. Professor Randall Davis of MIT summed it up at the National Research Counsel in 1990:
>> There is almost no way to visualize software. Sure, we have flow charts, we have data-flow diagrams, we have control flow diagrams, and everybody knows how basically useless those are. Flow charts are documentation you write afterward -- because management requires them, not because they are a useful tool.
> A patent is most similar to a real property deed specifying the metes and bounds for a parcel of land. Both documents are not easily understood but succeed if they secure the owners' interests in the specified claims. If the goal is to inform the world of an invention, software professionals have avenues more timely and less expensive than pursuing a patent application. In fostering the trade-off between the interests of inventors and the public, the source code is the best way to explain an algorithm.

The court's reasoning in the enablement and best mode cases has another implication as well. Because the court views actually writing and debugging a program as a "mere clerical function" "within the skill of the art," it follows that the court is unlikely to consider the work of programming itself to be sufficiently innovative to meet the nonobviousness threshold of section 103. After all, the same tests for adequacy of disclosure – would one of ordinary skill in the art be able to make the patented invention without undue experimentation – are also central to the obviousness inquiry.[40]

While only a limited number of appellate decisions discuss obviousness in the context of software patents, there is some reason to believe the court is imposing a rather strict standard. The first case involving the obviousness of a software-implemented invention is, perhaps surprisingly, a Supreme Court case from the 1970s. In *Dann v. Johnston*,[41] the Court held a patent on a "machine system for automatic record-keeping of bank checks and deposits" invalid for obviousness. The Court took a rather broad view of obviousness in the computer industry, focusing on whether analogous systems to the patentee's had been implemented in computer before, rather than analyzing the precise differences between the patentee's program and the prior art programs. The clear implication of the opinion is that if a reasonably skilled programmer could produce a

---

       Under this proposal, a computer system's complete source code would not have to be appendixed to the patent. The applicant would only have to include the source code directly relevant to enabling the claim language. In cases where claims are broadly written (as in a means-plus-function apparatus claim that covers the automation of an entire industry), a nearly complete program listing would be required.

[40]   *Compare* In re Vaeck, 947 F.2d 488 (Fed. Cir. 1991) (levels of experimentation and skill in the art in obviousness test) *with* In re Wands, 858 F.2d 731 (Fed. Cir. 1988) (levels of experimentation and skill in the art in enablement test). *See also* Donald S. Chisum, *Anticipation, Enablement and Obviousness: An Eternal Golden Braid*, 15 **AIPLA Q.J.** 57 (1987) (discussing the fundamentally interrelated nature of the obviousness and enablement inquiries).

[41]   425 U.S. 219 (1976).

program analogous to the patented one, and if there was motivation in the prior art to do so, the patented program is obvious.

The Federal Circuit has found software patents invalid for obviousness in two recent cases, *Lockwood v. American Airlines*[42] and *Amazon.com v. Barnes & Noble*.[43] Neither case opined directly on the ease with which computer programs could be produced, but both read obviousness as a rather substantial hurdle to patentability of software.[44] In *Lockwood*, the question was whether the defendant's own system made the patented claims obvious. The system had been in public use, but American Airlines had kept the workings of the system secret. Nonetheless, because Lockwood's patent was claimed in broad functional terms, the court found that similarly broad functional disclosures in the prior art were sufficient to render the patent obvious. While Lockwood argued that the information provided wasn't sufficient for one skilled in the art to make and use the system, the court pointed out that it was as detailed as the information Lockwood's own patent provided.[45] Thus, the patent's meager disclosure of technical details indirectly contributed to the court's finding of obviousness. In *Amazon.com*, the court found Amazon's "one-click" shopping feature to be obvious in view of certain references describing the desirability or feasibility of such a system in general terms, and one prior system that delivered data in response to a mouse click. The court rejected arguments that the one-click feature was technically difficult to implement, relying on the fact that the prior art generally described such a system as both desirable and feasible. The court also gave surprisingly short shrift to Amazon's evidence of secondary considerations of nonobviousness.[46]

---

[42] 107 F.3d 1565 (Fed. Cir. 1997).

[43] 239 F.3d 1343 (Fed. Cir. 2001).

[44] In *In re Zurko*, 111 F.3d 887 (Fed. Cir. 1997), the Federal Circuit held that a patented software invention was nonobvious even though each of the elements of the invention could be found in the prior art, where the prior art did not identify the problem to be solved. While *Zurko* certainly demonstrates that some software patents will be held nonobvious, it is a specific holding of rather limited utility to most software patentees.

[45] 107 F.3d at 1570.

The likely result of the Federal Circuit's focus on high-level functionality is that improvements in programming techniques will be found non-obvious in view of prior art that solved the same basic problem in a somewhat different way. This was arguably the result in both *Dann* and *Lockwood*, and it seems to follow from the court's view in the section 112 cases that programmers are an extremely skilled bunch. While disclosure is a minimal hurdle for software patents, then, obviousness can be a rather tougher one.[47]

---

[46] *Amazon.com v. Barnesandnoble.com,* 239 F.3d 1343, 1366 (Fed. Cir. 2001). To be sure, the court may have treated Amazon's patent more harshly because the case arose on appeal from a preliminary injunction. The court suggested that preliminary injunctions should not be granted if there were any serious question as to the validity of the patent. *Id*. at 1350-51. Whether it would apply as strict a test of obviousness after trial is not clear.

[47] This assumes that courts have access to the prior art necessary to make a realistic obviousness determination. A number of commentators have expressed concern about the difficulty Examiners have in finding software prior art. As Julie Cohen explains:
> [I]n the field of computers and computer programs, much that qualifies as prior art lies outside the areas in which the PTO has traditionally looked -- previously issued patents and previous scholarly publications. Many new developments in computer programming are not documented in scholarly publications at all. Some are simply incorporated into products and placed on the market; others are discussed only in textbooks or user manuals that are not available to examiners on line. In an area that relies so heavily on published, "official" prior art, a rejection based on "common industry knowledge" that does not appear in the scholarly literature is unlikely. Particularly where the examiner lacks a computer science background, highly relevant prior art may simply be missed. In the case of the multimedia data retrieval patent granted to Compton's New Media,[47] industry criticism prompted the PTO to reexamine the patent and ultimately to reject it because it did not represent a novel and nonobvious advance over existing technology. However, it would be inefficient, and probably impracticable, to reexamine every computer program-related patent, and the PTO is unlikely to do so.

Julie E. Cohen, *Reverse Engineering and the Rise of Electronic Vigilantism: Intellectual Property Implications of "Lock-Out" Technologies*, 68 **S. Cal. L. Rev.** 1091, 1179 (1995). *See also* Cohen & Lemley, *supra* note __, at 42-44; Greg Aharonian, http://www.bustpatents.com. *Cf.* John R. Allison & Mark A. Lemley, *Who's Patenting What? An Empirical Exploration of Patent Prosecution*, 53 **Vand. L. Rev.** 2099, 2131-32 (2000) (software patents actually cite slightly more non-patent prior art than other types of patents do).

Most of this criticism has been directed at the failure of the PTO to find (and patent applicants to cite) the relevant prior art. But parties in litigation have more time and money to spend, and are much more Patent scope is necessarily interrelated with obviousness and enablement.[48] The breadth of patent protection is in part a function of how different the invention is from the prior art. Further, patent claims are invalid if they are not fully described and enabled by the patent specification, so the permissible breadth of a patent will be determined by how much information the court determines must be disclosed to enable one of ordinary skill in the art to make and use the patented invention. The scope of the doctrine of equivalents is also a function of obviousness and enablement, since a patentee is not permitted to capture ground under the doctrine of equivalents that it would not have been permitted to claim in the first place.[49]

The Federal Circuit's treatment of software validity issues suggests that while the court will find relatively few software patents nonobvious, those that it does approve will be entitled to broad protection. The evidence on software patent claim scope so far is mixed, though there is some evidence tending to support this hypothesis. Most notably, in *Interactive Gift Express v. Compuserve*,[50] the patentee had designed a kiosk system for printing copyrighted works on demand. The Federal Circuit held that the claims of the patent should be read broadly, to cover any form of online downloading in response to a remote request.[51] In doing so, it reversed the district court's construction of five separate claim elements. As construed by the Federal Circuit, the patent is breathtaking in its

---

likely to find the best prior art than the PTO. The likely result is that while numerous software patents will issue, a large number of those actually litigated will be found obvious.

[48] *See* Chisum, *supra* note __.

[49] *See* Wilson Sporting Goods v. David Geoffrey & Assoc., 904 F.2d 677 (Fed. Cir. 1990).

[50] __ F.3d __, 2001 WL 792669 (Fed. Cir. July 13, 2001).

[51] *Id*.

scope, and most electronic commerce sites that permit downloading of digital information are likely within its ambit.

The court's treatment of software patent scope under the doctrine of equivalents has been more mixed. Many of these decisions have rejected application of the doctrine of equivalents to read claim language written for one product generation at such a high level of abstraction that it covers accused products from a different generation. Thus, in *Alpex Computer Corp. v. Nintendo Co.*,[52] the Federal Circuit held that a patent claim to a video game output display system was not infringed by a next-generation system that worked in a different way. *Alpex*'s claimed system included a display RAM that stored information corresponding to each pixel of a television screen in a discrete location. Nintendo's accused device, by contrast, used shift registers to store one "slice" of the video display at any given time. The Federal Circuit rejected a jury finding that the two systems were equivalent.[53] In *Digital Biometrics, Inc. v. Identix, Inc.*,[54] the court construed narrowly a patent claim to "image arrays" storing a two-dimensional slice of video data, and which were merged into a "composite array" storing a fingerprint image. The court held that the defendant's systems, which constructed the composite array directly rather than by using two-dimensional slices, did not create "image arrays" within the meaning of the claims. Most recently, in *Wang Laboratories, Inc. v. America*

---

[52] 102 F.3d 1214 (Fed. Cir. 1996).

[53] *Id.* at 1222. To similar effect as *Alpex* is *Wiener v. NEC Elec., Inc.,* 102 F.3d 534 (Fed. Cir. 1996). In that case, the Federal Circuit upheld the district court's finding of noninfringement under the doctrine of equivalents, because there were substantial differences between the patent's requirement that a computer program "call on" columns of data one byte at a time and the defendant's product, in which the columns alleged to be equivalent were not in the data matrix, and therefore were not called upon to read data. The court rejected the "conclusory" declaration of plaintiff's expert that the two processes were identical.

[54] 149 F.3d 1335 (Fed. Cir. 1998).

*Online*,⁵⁵ the court affirmed a district court decision granting summary judgment of noninfringement under the doctrine of equivalents. The patent claims in that case covered "frames," defined in the specification as pages encoded in character-based protocols. The court rejected Wang's attempt to extend the patent to cover bit-mapped pages, crediting evidence that there were "huge, huge differences" between the two approaches.⁵⁶

Other cases have applied the doctrine of equivalents more broadly. In some of those cases, the Federal Circuit has found equivalence between two different types of software programs written in different product generations. More troubling, some cases suggest that software implementations of certain ideas are equivalent to older mechanical implementations. An example is *Overhead Door Corp. v. Chamberlain Group, Inc.*⁵⁷ The patented system claimed a (mechanical) switch connected to a microprocessor, which could store the codes of multiple garage doors. The Federal Circuit held that the claim was not literally infringed by an electronic switch implemented in software. However, the court reversed a grant of summary judgment to the defendants under the doctrine of equivalents, concluding that a reasonable jury could find that the difference between mechanical and software implementations was a mere "design choice." *WMS Gaming, Inc. v. International Game Technology*⁵⁸ is also instructive. In that case, the court held that a claim written in means-plus-function language that relied for its corresponding structure on a computer programmed with a particular algorithm was limited in literal scope to the particular algorithm chosen and equivalents thereof. However, the court found the defendant's algorithm infringing under the doctrine of equivalents. This latter approach has the potential to expand the scope of patents in the software industry dramatically.[59]

---

⁵⁵ 197 F.3d 1377 (Fed. Cir. 1999).

⁵⁶ *Id*. at 1386. In a related context (interpreting equivalent structure in a means-plus-function claim), the court held that Nintendo's video game systems did not infringe GE's television switch patents because the patents, written in means-plus-function format, did not disclose a function for the switches identical to Nintendo's function. *General Electric Co. v. Nintendo Co.*, 179 F.3d 1350 (Fed. Cir. 1999). On how the doctrine of equivalents differs from equivalence under a means-plus-function analysis, see *Chiuminatta Concrete Concepts, Inc. v. Cardinal Indus.*, 145 F.3d 1303 (Fed. Cir. 1998).

⁵⁷ 194 F.3d 1261 (Fed. Cir. 1999).

Software patents, then, are likely to face serious obviousness hurdles. The few patents that overcome those hurdles need disclose virtually nothing about the detailed workings of their invention, and will likely be broadly interpreted to cover a variety of mechanisms for implementing the basic software invention.

### C. Biotechnology Patent Cases

In contrast to the Federal Circuit decisions regarding software, recent decisions involving genetic material have imposed a stringent disclosure standard for patenting macromolecules. The Court has laid particular emphasis on the "written description" requirement of section 112, which requires the patentee to specifically describe the claimed invention as part of the disclosure. The justification for such a detailed description is to demonstrate to others of ordinary skill that the inventor in fact has the invention in her possession; the assumption being that a sufficiently detailed description would not be possible if the inventor were speculating or guessing about its features.[60]

---

[58] 184 F.3d 1339 (Fed. Cir. 1999).

[59] For an argument that a variety of structural tendencies are likely to drive the courts to read software patent claims broadly under the doctrine of equivalents, see Cohen & Lemley, *supra* note __, at 39-50.

[60] Of course, in the case of constructive reduction to practice, or filing a "paper patent" without having acutally made the invention, the inventor is in some sense speculating or guessing about the features of an invention not yet built. But even in that instance, the underlying assumption in patent law is that the

This requirement is separate from, and potentially more stringent than that of enablement. Although the two are closely connected, satisfying one requirement does not necessarily satisfy the other. The classic example offered by one court is the situation in which the description of a particular chemical compound enables one of ordinary skill to make other, related, compounds, yet those other compounds are not described in the patent disclosure. The first compound is both enabled and described; the others are only enabled.

This venerable chemical patenting hypothetical has been brought to life by the Federal Circuit's biotechnology opinions. For example, in *Fiers v. Ravel*, the court considered the decision of the Patent Office in a three-way interference over patent applications drawn to DNA coding for human fibroblast beta-interferon (β-IF). One of the applicants, Ravel, relied for priority upon his Isreali patent application, which disclosed methods for isolating a fragment of the DNA sequence coding for β-IF and for isolating messenger RNA coding for β-IF. The court considered, inter alia, whether the disclosure in Ravel's Isreali application satisfied the U.S. written description requirement so as to form the basis for a U.S. application. The Federal Circuit upheld a determination by the Board of Patent Appeals and Interferences that Revel's disclosure was not an adequate description, largely because it failed to disclose the nucleotide sequence for the DNA molecule at issue. According to the court's reasoning, disclosing a method for obtaining the molecule was not the same as disclosing the molecule itself:

> An adequate written description of a DNA requires more than a mere statement that it is part of the invention and a reference to a potential method for isolating it; what is required is a description of the DNA itself. . . . A bare reference to a DNA with a statement that it can be obtained by

---

inventor "has" the invention mentally, and so can give a sufficiently detailed description of that inventive conception – physically creating the invention is mere

> reverse transcription is not a description; it does not indicate that Revel was in possession of the DNA.[61]

Since the Ravel application did not disclose the sequence for the molecule claimed, the court characterized it as disclosing merely "a wish, or arguably a plan, for obtaining the DNA."[62] A similar conclusion was reached in a subsequent *Eli Lilly v. Genentech*, where the patent at issue was drawn to a microorganism carrying the DNA sequence coding for human insulin, and supported this claim by disclosing a method for obtaining the human cDNA, as well as the amino acid sequences for the insulin protein. But relying on the *Fiers* opinion, the court concluded that the written description requirement again was not met: "Describing a method of preparing a cDNA or even describing the protein the cDNA encodes, as the example does, does not necessarily describe the DNA iteslf."[63]

In reaching these holdings, the Federal Circuit has been adamant that the degree of specificity required for an adequate description of nucleic acids requires "structure, formula, chemical name, or physical properties."[64] In *Eli Lilly*, because "[n]o sequence information indicating which nucleotides constitute human cDNA appears in the patent . . . . the specification does not provide a written description of the invention . . . ."[65] The court in such cases seems particularly incensed by applicants who designate a macromolecule by English nomenclature, such as "vertebrate insulin cDNA":

> A definition by function . . . is only an indication of what a gene does, rather than what it is. It is only a definition of a useful result rather than a definition of what achieves that

---

[61] 984 F.2d at 1170-71.
[62] *Id.*
[63] 119 F.3d at 1567.
[64] 984 F.2d at 1170
[65] 119 F.3d at 1567.

> result. Many such genes may achieve that result. The description requirement of the patent statute requires a description of an invention, not an indication of a result that one might achieve if one made that invention. Accordingly, naming a type of material generally known to exist, in the absence of knowledge as to what that material consists of, is not a description of that material.[66]

Such failure to describe more than one or two nucleotides is a particular problem where the patent claims are drawn to a broad class of nucleotides. For example, Revel's claim covered all DNA molecules that code for β-IF, but: "[c]laiming all DNAs that achieve a result without defining what means will do so is not in compliance with the description requirement; it is an attempt to preempt the future before it has arrived."[67] The construction of the written description requirement as requiring precise sequence data gains particular significance whenever claims are drawn to an entire genus, or family, of molecules. The patent discussed in the *Lilly* written description analysis claimed a broad family of DNA molecules coding for insulin in different mammalian species, but it disclosed only one species of DNA, that coding for rat insulin. The court held this to be insufficient to describe the broad class of cDNAs coding for mammalian or vertebrate insulin. Although declining to specify exactly what would be needed to support a broad claim, the court cited previous chemical cases dealing with related groups of small molecules. Based on these cases, the court declared that macromolecules should be treated in the same fashion; the patentee need not not show every member of a claimed genus, but is required to show a "representative" number of cDNAs, illustrating or defining the common structural features of a "substantial" portion of the genus.

---

[66] *Id.*
[67] at 1171

A similarly broad claim was rejected in the *Amgen* case as failing the standard for enablement.[68] There, the patentee claimed nucleic acid sequences coding for the protein erythropoetin, or for other proteins with the same biological function. The trial judge concluded that because Amgen was unable to specify which of these analogs might have the biological properties claimed, the claims were not enabled. The Federal Circuit panel, however, that the district court had reached the right conclusion for the wrong reason: Whereas the district court focused on the thousands of EPO analogs that could be created by substituting amino acid residues in the polypeptide chain, the appellate court focused on the patentee's failure to disclose the DNA molecules that would code for those analogs. Since the claims were directed to DNA sequences, the issue was not the enablement of the EPO analogs, but rather the enablement of the myriad DNA sequences, which the court held could not be made and used on the basis of a few examples.

The same concerns that characterize the Federal Circuit's jurisprudence of biotechnology disclosure – the inadequacy of methodological disclosure, the requirement to specify sequence or structure, and uncertainty of selection within large classes of homologous molecules -- have shaped the Federal Circuit's jurisprudence of biotechnology. However, in the case of obviousness, the issue has been the presence of such factors in the prior art, rather than in inventor's disclosure. Thus, the Federal Circuit held in *In re Bell* that a claim to DNA coding for human insulin-like growth factor (hIGF) was not obvious although the prior art disclosed the amino acid sequence for the hIGF proteins and a method for using that information to obtain the corresponding DNA molecule.[69] Under similar facts in *In re Deuel*, the court found claims directed to DNA coding for heparin binding growth factors (HBGFs) were not obvious in light of prior art disclosure of a partial amino acid sequences and a method for using that information to obtain the corresponding DNA molecule.[70]

---

[68] 927 F.2d 1200, 1212 (Fed. Cir. 1991).
[69] 991 F.2d 781 (Fed. Cir. 1993).

Each decision rested largely upon the court's perception that the actual sequence of the claimed DNA molecules was uncertain or unpredictable from the prior art. In both cases the court dismissed as irrelevant the biological relationship between the molecules disclosed in the prior art and those claimed by the patent. The amino acid sequences of the proteins disclosed in the prior art are ultimately determined by the sequence of RNA nucleotides coding for the protein, which is in turn determinative of the cDNA claimed in the patent.[71] The correspondence of nucleotide sequences to amino acid sequences is well known as key to the "central dogma" of molecular biology: the transfer of genetic information from DNA to RNA to protein chains. However, certain amino acids can correspond to more than one nucleotide sequence, introducing uncertainty into the inverse relationship: that of amino acid sequence to nucleotide sequence. Because of this redundancy or "degeneracy" in the genetic code, the court noted in *Bell* that a vast number of possible sequences – about $10^{36}$ – might code for the protein sequences disclosed in the prior art. The plaintiff claimed only one of these, in essence having searched among a large number of possibilities to select the particular sequence coding for hIGF.

Numerous commentators have pointed out that such a search is relatively routine. But prior art disclosure of method, even admittedly obvious method, was held insufficient

---

[70] 51 F.2d 1552 (Fed. Cir. 1995).
[71] Neither *In re Bell* nor *In re Deuel* dealt with genomic DNA (gDNA) sequences, which are transcribed by cellular proteins to produce a messenger RNA molecule. Both cases considered non-naturally occurring cDNA sequences, which are reverse transcribed from messenger RNAs. The correspondence between

to cure such uncertainty of structure. In rejecting the DNA claims in *Bell* and *Deuel*, "the PTO's focus on known methods for potentially isolating the claimed DNA molecules is also misplaced because the claims at issue define compounds, not methods."[72] Prior to *Bell*, the opinion in *Amgen* had stressed the uncertainty of the methods for gene location available at the time of invention: "[I]t might have been feasible, perhaps obvious to try, to successfully probe a human gDNA library with a monkey cDNA probe, it does not not indicated that the gene could have been identified and isolated with any success. . . . there was no reasonable expectation of success in obtaining the EPO gene by the method that Lin eventually used."[73] By the time of the research at issue in *Bell*, such methods for searching a large universe of molecules were perhaps painstaking and time-consuming, but had an established likelihood of success.

Yet the court defined the issue in *Bell* and *Deuel* not as a matter of the uncertainty of obtaining a particular sequence, but of the uncertainty of predicting or visualizing from the prior art what sequence would be found. Even in the *Amgen* opinion, the court hinted that the key to obviousness lay in the prediction of an exact sequence, as "[n]either the DNA nucleotide sequence nor its exact degree of homology with the monkey EPO gene was known at the time."[74] And in *Deuel*, the court explicitly held that "until the claimed molecules were actually isolated and purified, it would have been highly unlikely for one of ordinary skill in the art to contemplate what was ultimately obtained. *What cannot be contemplated or conceived cannot be obvious*."[75] Thus a likelihood, or even a certainty

---

gDNA and RNA may be very different than than that of cDNA to RNA, especially in eukaryotic organisms where the processing of RNA transcripts may be extensive.
[72] 51 F.3d at 1558; *see also* 991 F.2d at 785 (same).
[73] 927 F.2d at 1208.
[74] 927 F.2d at 1208-09.
[75] 51 F.3d at 1558 (emphasis added).

of finding a DNA molecule with particular properties was deemed essentially irrelevant to whether structural claims to that molecule are obvious.

The corollary to this holding is that a molecule will be obvious if the sequence, and not particular function, are discernible in the prior art. Prior art description of the "general idea of the claimed molecules, their function, and their general chemical nature"[76] is insufficient to render a molecule obvious. Some commentators have suggested that this requirement for obviousness stands some danger of collapsing into the standard for anticipation: under section 102 of the Patent Act, an invention lacks patentable novelty if is elements are fully described in a prior art reference, and the Federal Circuit's obviousness requirement could be read to require such a prior art anticipation as the effective standard for obviousness.[77] But unlike the requirements for anticipation, the Federal Circuit's biotechnology obviousness standard appears to require that the sequence of the DNA be predictable from the prior art, and not necessarily explicitly described. For example, the court in *Deuel* suggests that "a protein of sufficiently small size and simplicity, so that lacking redundancy, each possible DNA would be obvious over the protein."[78] Although the Federal Circuit has not explicitly held so, one would also suspect that disclosure in the prior art of a substantial number of homologous sequences would render a new homologue predictable, and so render it obvious – just as the court has held that disclosure of a substantial number of homologues enough to satisfy the written description requirement for that genus.

---

[76] 51 F.3d at 1558.
[77] Indeed, the Federal Circuit has several times suggested that the two patent standards are closely linked, characterizing obviousness as a sort of continuum with anticipation as the "epitome" or "ultimate" endpoint of obviousness. *See, e.g.*, Jones v. Hardy 727 F.2d 1524 (Fed. Cir. 1984); Connell v. Sears, Roebuck & Co. 722 F.2d 1542 (Fed. Cir. 1983); In re Baxter Travenol Labs, 952 F.2d 388 (Fed. Cir. 1991).
[78] 51 F.3d at 1559.

Note that these holdings are all of a piece with the court's earlier holdings, such as the rejection, on disclosure grounds, of Revel's claim to all DNA sequences coding for β-IF. Due to degeneracy in the genetic code, Revel could not adequately describe the claimed invention as DNA coding for β-IF; an astronomically large number of possible sequences might do so. And if a functional or narrative description in a patent is insufficient to properly describe a DNA molecule coding for β-IF, so the presence of a functional or narrative description of β-IF protein in the prior art would be insufficient to render the molecule obvious: according to the court, one cannot describe what one has not conceived, and what cannot be contemplated or conceived cannot be obvious. Just as disclosure in a patent of a method for obtaining a particular cDNA is inadequate to properly describe the invention, so disclosure in the prior art of a method for obtaining a particular cDNA cannot render the claimed invention obvious.

The conceptual linkage of obviousness and enablement to the depiction of macromolecular sequences in, respectively, the prior art or the patent disclosure, dictates a particular and predictable result for the availabilty and scope of such biotechnology patents. The expected outcome is that DNA patents will be extremely narrow but numerous. Under the Federal Circuit's precedent, a researcher will be able to claim only sequences disclosed under the stringent written description -- the actual sequence in hand, so to speak. At the same time, the inventor is shielded from obviousness by the lack of such explicit and detailed disclosure in the prior art. This lack of effective prior art seems to dictate that anyone who has isolated and characterized a DNA molecule is certain to receive a patent on it. But the inventor is certain to receive a patent only on that molecule, as the Federal Circuit appears to regard other related molecules as inadequately described until their is sequence is disclosed.

The set of axioms underlying this set of results forms a logical framework that can be predictably extended to other biotechnology inventions besides DNA. For example, one would conclude from the Federal Circuit's analysis that a cDNA should be obvious in light of its corresponding mRNA, since the former is reverse transcribed from the latter, and there is no redundancy or degeneracy in the correspondence between the nucleotides in the two molecules.[79] However, an mRNA or corresponding cDNA needd not render obvious the genomic DNA (gDNA) from which is is derived, since in many organisms, the gDNA will include intervening sequences, or introns, that are not predictable from the mRNA sequence.

Perhaps more important than the extension of the Federal Circuit's logic to other classes of molecules is the extension of its logic to other patent doctrines. For example, as we have indicated with regard to software, the infringement doctrine of equivalents is directly linked to the standard for obviousness: the patentee is prevented from capturing by equivalents subject matter that would have been obvious at the time he obtained his patent.[80] Given the very parsimonious reading that the Federal Circuit gives to biotechnology obviousness, there should be enormous latitude for findings of equivalence in nucleotide infringement cases. The overall result would seem to be an unsustainable configuration of patentability, in which large numbers of narrow patents, with broad scopes of equivalence, blanket biotechnology research.

---

[79] L. STRYER, BIOCHEMISTRY 132 (3d ed. 1988); J.WATSON et al., MOLECULAR BIOLOGY OF THE GENE 610-11 (4$^{th}$ ed. 1987).

II.     Explanation for the difference - PHOSITA

A.  The Role of the PHOSITA

Our survey of the biotechnology and the software patent cases highlights an important reciprocal relationship between obviousness and disclosure. In biotechnology, where highly detailed disclosure is required to satisfy the enablement and written description standards, similarly detailed disclosure is required to render the invention obvious. In software, where little specific detail is needed to satisfy the requirements of disclosure, similarly little detail is needed to render the invention obvious. In each case, the court takes the patentability requirements of non-obviousness and disclosure as firmly tied to a common standard.

The common standard connecting these requirements is that under each statutory section, the respective requirements must be addressed with regard to the "person having ordinary skill in the art," sometimes known by the acronym of PHOSITA.[81] Much of the caselaw concerning the PHOSITA arises out of the consideration of the obviousness standard found in section 103 of the patent statute.[82] Although originally developed as a common law doctrine, the non-obviousness criterion was codified in the 1956 Patent Act as a requirement that the claimed invention, taken as a whole not be obvious to one of ordinary skill in the art at the time the invention was made.

---

[80] Wilson Sporting Goods Co. v. David Geoffery & Ass., 904 F.2d 677 (Fed. Cir. 1990) (testing equivalence by inquiring whether a hypthetical claim encompassing the accused product would have been obvious at the time of invention).
[81] John O. Tresansky, PHOSITA -- The Ubiquitous and Enigmatic Person in Patent Law, 73 J. Pat. & Trademark Off. Soc'y 37 (1991); see generally R. HARMON, PATENTS AND THE FEDERAL CIRCUIT § 4.3 (5th ed. 2001)
[82] 35 U.S.C. § 103.

The PHOSITA is equally central to calibrating the legal standard for patent disclosure. As the quid pro quo for her period of exclusive rights over an invention, the inventor must fully disclose the invention to the public. The first paragraph of 35 U.S.C. sec. 112 requires that this disclosure enable "any person skilled in the art" in the art to make and use the claimed invention. The language this section therefore indicates that the inventor's compliance with the requirement of enablement should be measured with reference to a standard similar or identical to that in section 103; indeed, the language appears to tie the enablement requirement to non-obviousness via this shared metric.

This same language set the metric for several related disclosure doctrines as well. First, the definition of enablement effects the patentablity requirement of specific utility, as the invention must operate as described in the specification if the inventor is to enabled one of ordinary skill to use it. Additionally, compliance with the independent requirements of description and best mode disclosure are measured with reference to the understanding of a "person skilled in the art." And finally, the definiteness of patent claims, which must be written so as to warn members of the public just what is and is not covered by the patent, are assessed with regard to the knowledge of one having ordinary skill in the art -- if the terms of the claims would not be comprehensible to such a person, then they fail the requirements of section 112.

The PHOSITA is nothing if not versatile, and may also show up as a convenient metric in other unexpected areas, including judicially created patent doctrines. For example, the PHOSITA reappears in some formulations of the standard for infringement by equivalents. In its germinal opinion on the doctrine of equivalents, *Graver Tank*, the Supreme Court indicated that the equivalence between elements of an allegedly infringing device and those of a claimed invention might be tested by determining whether the such elements were known in the art to be substitutes for one another.[83] A great deal of patent law thererfore rests upon the measurement of some legal parameter against the skill and knowledge of the PHOSITA.

This is not to say the PHOSITA has any actual skill or knowledge. Like her cousin, the reasonably prudent person,[84] the PHOSITA is something of a juridical doppelganger,[85] embodying a legal standard for patentability rather than the actual capability of any individual or group of individuals. Courts have on occasion equated the knowledge of a given individual, such as a patent examiner, with that of the PHOSITA. But fine line between taking the skill of an examiner or other artisan as probative evidence of the level of skill in the art, and equating the skill of such persons with the characteristics of the hypothetical PHOSITA.

This places the standard for patentablity on a legally objective, rather than subjective, footing. The PHOSITA standard measures the inventor's achievements against a judicially determined external metric, rather than against an expectation based on whatever level of skill the inventor might actually possess. The standard also has the practical effect, since the question is one of law, of avoiding the requirement that judges and other arbiters of patentability be experts in a given field. The PHOSITA standard is thus an ultimate conclusion of law based upon evidence,[86] but not dictated by the capabilities or knowledge of the Patent Office examiner, a reviewing judge, or even that of the inventor:

> Realistically, courts never have judged patentablity by what the real inventor/applicant/patentee could or would do. Real inventors, as a class, vary in their capacities from ignorant geniuses to Nobel Laureates; the courts have always applied a standard based on an imaginary worker of their own devising whom they have equated with the inventor.[87]

---

[83] Graver Tank Mfg. Co. v. Linde Air Products Co. 339 U.S. 605 (S. Ct. 1950)
[84] *See* Panduit Corp. v. Dennison Mfg. Co. 810 F.2d 1561 (Fed. Cir. 1987) (comparing the PHOSITA to the "reasonable man.")
[85] *Id.* (characterizing the PHOSITA as a "ghost.").

The standard is thus objective in the sense that it does not inquire into a particular inventor or artisan's level of skill. But this does not mean that it is static or fixed. This definition offers courts a number of consitutent factors that may be adjusted to modulate the requirements for patentability under different circumstances. The first of these is the definition of the particular "art" in which the PHOSITA is deemed to have ordinary skill. The PHOSITA is generally portrayed as having comprehensive knowledge of the references in the particular art. But the parameters of the art are subject to fluctuation, and thus so is the size and depth of the library of references with which the PHOSITA is presumed to be familiar. For example, in the case of a DNA patent, is the relevant art biochemistry or molecular biology, or biology in general? Courts have attempted to avoid drawing such boundaries by defining the PHOSITA's knowledge as that reasonably pertinent to the problem the inventor was trying to solve. But this requires that the court engage in the equally mercurial exercise of defining of the problem that the inventor had under consideration.

---

[86] Panduit Corp. v. Dennison Mfg Co., 810 F.2d 1561 (Fed. Cir. 1987).

A second circumstantial variable is the level of skill that would be considered "ordinary." Unlike the inventor, who almost by definition is presumed to be one of extraordinary skill,[88] the PHOSITA standard contemplates some median level of skill. In assessing that median level, courts may take into account a long list of factors, including the approaches found in the prior art, the sophistication of the technology involved, the rapidity of innovation in that field, and the level of education typical of those in the field.[89] The courts have also endowed the PHOSITA with mediocre personality traits; she is conceived of as an entity that adopts conventional approaches to problem solving, and is not inclined to innovate, either via exceptional insight or painstaking labor.[90]

Some care must be exercised in characterizing the PHOSITA, as it is tempting to do so on the basis of an unfounded presumption, which is that the PHOSITA remains constant from section to section of the patent statute. On the contrary, some commentators have recognized as quite possible the possibility that the imaginary artisan found in these different statutory sections, though bearing the same denomination, might well display different and even inconsistent characteristics as between the different sections.[91] The PHOSITA for purposes of obviousness may not necessarily be the PHOSITA for purposes of enablement, written description, definiteness, or equivalence. Because she is a legal construct designated to embody certain legal standards, the PHOSITA could well change depending on the purpose she is serving at the time.

---

[87] Kimberly-Clark Corp. v. Johnson & Johnson, 745 F.2d 1437 (Fed. Cir. 1984). *See* Hodosh v. Block Drug Co., 786 F.2d 1136 (Fed. Cir. 1986) (actual inventors cannot be required to have the omniscience of the figurative person of ordinary skill); In re Nilssen

[88] Standard Oil Co. v. American Cyanamid Co., 774 F.2d 448 (Fed. Cir. 1995).

[89] See, e.g., Bausch & Lomb Inc. v. Barnes-Hind, Inc., 796 F.2d 443 (Fed. Cir. 1986) (listing pertinent factors); see also Helifix Ltd. v. Blok-Lok Ltd, 208 F.3d 1339 (Fed. Cir. 2000) (district court erred by failing to consider these factors).

[90] Standard Oil Co. V. American Cyanamid Co., 774 F.2d 448 (Fed. Cir. 1985)

[91] *See* Tresansky, *supra* note __ at 52-53.

Perhaps because we have the most caselaw discussing the PHOSTIA in the contexts of obviousness and of enablement, some dispararity in fact appears in the judicial characterization of the PHOSITA for purposes of these issues. The section 103 PHOSITA appears to be something of a problem solver, who the courts set to work hypothetically solving the problem solved by the inventor.[92] To be sure, the obviousness PHOSITA is not an especially inspired problem solver, as he is imagined to remain stuck in the rut of conventional thinking.[93] By contrast, the PHOSITA of the first paragraph of section 112 shows no such innovative tendency, but is simply a user of the technology. If the enablement PHOSITA shows any problem solving ability, it is in tapping the prior art to fill in gaps left by the inventor's disclosure – a rather different skill than that of the obviousness PHOSITA.[94] The two PHOSITAs also differ in the knowledge imputed to them. The knowledge of one is assessed as of the time of invention, but that of the other at the time a patent is filed -- due to the passage of time, the latter universe of references is likely to be larger. But conversely, hidden or non-public references which may serve as prior art under section 103 are not necessarily imputed to the knowledge of the PHOSITA who make or use the invention under section 112, as such references are not readily available to the public.[95]

---

[92] *See* Orthopedic Equip. Co. v. United States, 702 F.2d 1005 (Fed. Cir. 1983); In re Grout, 377 F.2d 1019 (C.C.P.A. 1967).
[93] Standard Oil Co. v. American Cyanamid Co. 774 F.2d 448 (Fed. Cir. 1985).
[94] *See* Tresansky, *supra* note __ at 54.
[95] Quaker City Gear Works, Inc. v. Skil Corp. 747 F.2d 1446 (Fed. Cir. 1984); In re Howarth, 654 F.2d 103 (C.C.P.A. 1981)

B. Implication of PHOSITA approach

Patent law ostensibly offers a single set of standards, which are applied uniformly to all patentable subject matter, "anything under the sun made by man." Yet in another sense, every technology is treated individually. Within the broad confines of patent law, the PHOSITA standard offers courts the ability to adapt patent requirements to different technologies. In this sense, patent law is technology specific, in essence offering different and fact sensitive standards of disclosure and obviouness to different technologies.

Yet, recognizing that the PHOSITA standard dictates that different technologies will be accomodated in different ways, the developments that we have described in software and biotechnology seem to us extraordinary, and well beyond the accomodation offered by the PHOSITA standard. Consider, for example, the extremely stringent disclosure standard developed in the biotechnology cases. Such disclosure would be dictated under the PHOSITA analysis in order to assure those of ordinary skill that the inventor posessed the invention at the time the application was filed -- apparently, the Federal Circuit believes that biotechnology researchers need a very high degree of assurance. Computer programmers, on the other hand, apparently require very little assurance -- simply an indication of function will do. The opposite is true in each case with regard to obviousness; the court appears to believe that computer programmers can fully envision working code from only a suggestion of function, whereas biotechnologists apparently need genetic sequences explicitly spelled out in the prior art to render a molecule obvious.

Indeed, we wonder about the extension of such doctrine to other technological areas. For example, small-molecule chemistry has long had its own discrete set of patentability doctrines, developed in a long line of cases that attempt to accommodate the level of skill in that particular technology.[96] The rules articulated in this line of cases represents something of a compromise compromise between the predictable similarities in the characteristics of molecular families and the difficulty in predicting the effect of structure in three dimensions. As a first approximation, structural relatedness between molecules disclosed in the prior art and a novel molecule claimed in a patent gives rise to a prima facie case of obviousness.[97] However, chemical structures depicted two-dimensionally on paper may not accurately reflect the properties of a physical structure that exists in three dimensions -- molecules react with one another in three dimensions, and the three dimensional configuration dictates the chemical characteristics of the molecule. Thus, even in small molecules, the three-dimensional complexity arising from what appears on paper to be slight changes in structure may give rise to radically different properties in apparently related molecules. Even with three dimensional modeling, the effects of such complexity have long been difficult to predict. Such unpredicted characteristics occurred with enough frequency that a rule developed allowing a *prima facie* case of obviousness in small molecules to be rebutted by evidence of unpredictable or unexpected properties in the claimed molecule.[98] The technological assumption built into such a rule appears to be that the PHOSITA in small-molecule chemistry can generally predict the properties of a chemcial or group of chemcials, or may occasionally be surprised by their properties, but either outcome is based on the molecules' structural depiction.

---

[96] *See* In re Dillon, 919 F.2d 688 (1990) (recounting history of chemical obviousness cases).
[97] H. WEGNERr, CHEMICAL PATENT PRACTICE (1991).
[98] In re Papesch, 315 F.2d 381 (C.C.P.A. 1963); see also Harold Wegner, Prima Facie Obviousness of Chemical Compounds, 65 Am. Pat. L. Ass'n Q.J. 271 (1978).

The rule in these small molecule cases appears closely related to that announced in biotechnology cases discussed here -- the Federal Circuit has declared that DNA "is a chemical, albeit a complex one," and has articulated a desire to treat the patenting of macromolecules in the same fashion as the patenting of more traditional organic molecules. In focusing upon structural depiction as the linchpin of both obviousness and disclosure, the biotechnology cases rely upon, and appear to extend, the line of chemical cases summarized above. But just as we question the application of these rules to macromolecules, we are similarly uncertain that these special rules for obviousness in small molecule chemical cases are well suited to accommodate current chemical research practice, particularly in light of the rules articulated by the Federal Circuit for macromolecules.

In particular, modern techniques of rational drug design and combinatorial chemistry seem to push against this traditional construction of chemical obviousness in much the same way that the routinization of DNA probing pushes against the rules of patentability in the biotechnology cases. For example, small-molecule chemists now search for useful compounds by first specifying the functions that they hope to find.[99] The characteristics of desirable molecules are represented mathematically, in equations depicting functionally equivalent chemical groups and side chains.[100] Based on the predictions of such mathematical models, chemists can then search through large panels of related molecules, selecting those with the closest match to predicted function.[101]

---

[99] See generally Hugo Kubinyi, The Quantitative Analysis of Structure-Activity Relationships in 1 Burger's Medicinal Chemistry and Drug Discovery 497-571 (Manfred E. Wolff ed. 1995).
[100] Richard B. Silverman, The Organic Chemistry of Drug Design and Drug Action 26-34 (1992) (describing the Hansch equation that correlates biological activeing with physicochemical properties of drug candidates).

This methodology closely parallels the type of molecular "search" considered in most of the Federal Circuit macromolecule cases, where large libraries of DNA molecules are probed in order to identify those that correspond to an expected functional characterisitc -- i.e., the propensity to hybridize with probes of a particular nucleotide configuration, and concomitantly the capacity to code for cellular production of particular gene products.[102] Combinatorial chemistry, much like DNA probing, tends to focus upon the function of the end product, removing much of the uncertainty from the outcome of a search for a desired molecule. but not necessarily from predicting the precise structure of the molecule that is ultimately found. Indeed, role of chemical structure is to some extent marginalized, as dissimilar structures with similar functions may be treated as equivalent in narrowing the search.

### D.    The PHOSITA Standard Has Been Applied Inappropriately in the Software and Biotechnology Industries

Despite the anomalies that we have identified in the cases of biotechnology and of software, we continue to believe that the PHOSITA approach in general represents the proper standard for patent law. Basing the proof required on the level of skill in the art makes logical sense. At the simplest level, this approach patent is intended to benefit the public; people who work in a given technology must understand the patent as it relates the prior art, so it makes sense to take into account what that person knows in order to decide whether a patent is obvious or has been enabled. From a policy standpoint, the practicality of working in different technologies requires a flexible approach to determining disclosure or obviousness, and the PHOSITA approach gives a court that flexibility.

---

[101] *See* Dinesh V. Patel et al., *Applications of Small-Molecule Combinatorial Chemistry to Drug Discovery*, 1 Drug Discovery Today 134 (1996); Jan J. Scicinski, *Chemical Libraries in Drug Discovery*, 134 Trends in Biotechnology 246 (1995); Joseph C. Hogan Jr., *Directed Combinatorial Chemistry*, 384 Nature 17S (1996).
[102] *See generally* J. WATSON et al., RECOMBINANT DNA 104-07 (2d ed. 1991) (describing techniques for probing libraries of cloned genes).

Consequently, in order to identify the source of the anomalies in biotechnology and software, we look first to the Federal Circuit's application of this standard, rather than to the standard itself. One possibility, which has occurred to previous commentators as well as to us, is that the Federal Circuit application of the PHOSITA standard in these technologies is wrong as a matter of science. One reading of these cases is that the Federal Circuit seems to have substituted caricature for a nuanced understanding of the technology. The court has repeatedly suggested that programming itself is a "mere clerical function" that presumably does not warrant the grant of a patent. The court seems to consider only the ideas or functions of a computer program worth of patent protection. In short, the court thinks of programmers as of extraordinary skill, capable of implementing any idea in a computer program as a matter of course. Sometimes this assumption benefits patentees, notably in enablement and best mode determinations.[103] Other times, notably in obviousness cases, the assumption that programmers are extremely skilled works against the patentee.[104] But as a matter of computer science, the court's assumptions are just wrong. Any programmer will tell you that coming up with an idea for a computer program is rather less than half the battle.[105] Programs can take years to write even under the best circumstances. Some will simply not work. Others will require innovative programming techniques. Even once they are written, most programs have bugs that must we worked out in order for the program to be stable.[106] And in many cases, the process of writing the program changes the idea itself in a sort of continuous feedback loop.[107] Not only is it wrong to say that writing a program is a "mere clerical function" to a skilled programmer, but in fact many of the truly innovative improvements in computer software occur at the level of programming, not the idea to have a computer perform a specific function.[108]

---

[103] *See supra* notes __-__ and accompanying text (noting the low standards applied to software patents under section 112).

[104] *See supra* notes __-__ and accompanying text.

In the biotechnology cases, problem is the opposite; the court focuses repeatedly on the "uncertainty" inherent in the field, scoffing at claims drawn to molecular function, and demanding precise disclosure of any embodiment.[109] The court seems to believe that

---

[105] A good judicial discussion of this is *Computer Assoc. v. Altai, Inc.*, 982 F.3d 693 (2d Cir. 1992). *See also* 3 **Nimmer on Copyright** §13.03[F]; Peter S. Menell, *An Analysis of the Scope of Copyright Protection for Application Programs*, 41 **Stan. L. Rev.** 1045 (1989); Pamela Samuelson et al., *A Manifesto Concerning the Legal Protection of Computer Programs*, 94 **Colum. L. Rev.** 2308 (1994); Thomas M. Gage, Note, *Whelan Assoc. v. Jaslow Dental Laboratories: Copyright Protection for Computer Software Structure – What's the Purpose?*, 1987 **Wisc. L. Rev.** 859. *Cf.* Mark A. Lemley & David W. O'Brien, *Encouraging Software Reuse*, 49 **Stan. L. Rev.** 255, 261-66 (1997) (detailing the cost savings available from reusing computer code rather than reinventing it from scratch).

[106] *See* Lemley & O'Brien, *supra* note __, at 261-64 and sources cited therein.

[107] *See, e.g.,* Menell, *supra* note __, at __.

[108] *See* Mark A. Lemley & David W. O'Brien, *Encouraging Software Reuse*, 49 **Stan. L. Rev.** 255, 302 (1997) (encouraging protection for such ideas in preference to the more common patents on old ideas implemented in digital format). Indeed, Richard Stern and Julie Cohen have both proposed that software patents be limited to innovative programs. *See* Julie E. Cohen, Julie E. Cohen, *Reverse Engineering and the Rise of Electronic Vigilantism: Intellectual Property Implications of "Lock-Out" Technologies*, 68 **S. Cal. L. Rev.** 1091, 1169 (1995); Richard H. Stern, *Tales From the Algorithm War: Benson to Iwahashi, It's Déjà vu All Over Again*, 18 **AIPLA Q.J.** 371, 395 (1991).

[109] *See supra* notes __-__ and accompanying text (discussing the role of uncertainty in the Federal Circuit's biotechnology jurisprudence).

*Is Patent Law Technology-Specific?* Burk & Lemley                                                                              DRAFTbiotechnology is as much a black art as a science. In *Bell* and *Deuel* the court's belief in uncertainty benefits the patentee, since it means that knowledge of a protein and a method for deriving the cDNA sequence did not render the cDNA sequence obvious.[110] By contrast, the same assumption about uncertainty hurts patentees in cases like *Enzo, Lilly* and *Amgen*, because it precludes them from claiming any DNA sequence they have not actually described in the patent specification.[111] But all of these holdings are based on the assumption that one ordinarily skilled in biotechnology cannot get from a protein to a DNA sequence, or from the DNA sequence of one mammal to the corresponding DNA sequence of another mammal.

Arguably this understanding of the science of biotechnology is simply wrong.[112] Robert Hodges has argued that "the key event is the cloning of the first gene in a family of corresponding genes. Once a researcher accomplishes this very difficult task, the researcher can typically obtain other members of the gene family with much less effort."[113] Indeed, the process is largely automated. Such research is properly compared

---

[110] *See* In re Bell, 991 F.2d 781 (Fed. Cir. 1993); In re Deuel, 51 F.3d 1552, 1559 (Fed. Cir. 1995). *Cf. Fiers v. Revel*, 984 F.2d 1164 (Fed. Cir. 1993) (using the same standard in an interference proceeding to benefit one applicant at the expense of another). *But cf.* In re Mayne, 104 F.3d 1339 (Fed. Cir. 1997) (DNA sequence in prior art rendered obvious a claim to an altered version of that sequence that changed only one amino acid).

[111] *See* Enzo Biochem v. Calgene, Inc., 188 F.3d 1362, 1371 (Fed. Cir. 1999); Regents of the University of California v. Eli Lilly & Co., 119 F.3d 1559 (Fed. Cir. 1997); Amgen Inc. v. Chugai Pharmaceutical Co., 927 F.2d 1200 (Fed. Cir. 1991); In re Goodman, 11 F.3d 1046, 1052 (Fed. Cir. 1993).

[112] We acknowledge Lawrence Sung's contrary view, that the biotechnology cases are simply decided on their individual facts and do not reflect any patterns. *See* Sung, *supra* note __, at 107; John W. Schlicher, *Biotechnology and the Patent System: Patent Law and Procedures for Biotechnology, Health Care and Other Industries*, 4 **Balt. Intell. Prop. L.J.** 121, 127 (1996) ("I do not understand the Court of Appeals for the Federal Circuit to have created a subset of patent law doctrines for biotechnology."). But the cases seem to us much more consistently to depart from the standards applied in other industries.

[113] Robert A Hodges, *Black Box Biotech Inventions: When a Mere "Wish or Plan" Should be Considered an Adequate Description of the Invention*, 17 **Ga. St. U. L. Rev.** 831, 832 (2001). Also, see John M. Lucas, *The Doctrine of Simultaneous Conception and Reduction to Practice in Biotechnology: A Double Standard*

to searching a "black box" in which are contained molecules of known characteristics, if unknown structure -- the search is conducted on the basis of what is known, that is, the function, rather than on the basis of what is unknown -- the precise structure.  Thus, the outcome of such research is predictable with regard to the expected function of the molecule that will be found, and the expected success of finding such a molecule.

This explanation of the Federal Circuit's jurisprudence in these areas is not altogether satisfactory, as in biotechnology at least, it fails to explain the court's indifference to the technology subsequent to *Amgen*.  The obviousness decision in *Amgen* clearly rested upon the uncertain likelihood of success in the particular probing methodology used to find the EPO gene.  Had the court adhered to this analysis in later cases, carrying forward into subsequent opinions a static impression of biotechnological techniques, the poor fit between patent doctrine and patent policy could be easily explained.  But in those later cases, the court seems quite indifferent to the certainty or uncertainty of methodological success, fashioning instead a standard based on structural precision and foreseeability that ignores the state of technology, past or present.

This observation might be accommodated by an alternative explanation: that the court, rather than stumbling in its application of law to changing technology, is as a matter of law simply creating a unique enclave of patent doctrine for biotechnology.  The same explanation might be applied to the software cases: that it is not a unique set of facts applied to the PHOSITA construct that is generating a technology-specific body of patent law, but rather a deliberate manipulation of doctrine itself.  Yet this alternative explanation seems to us even less satisfactory than the first -- it essentially moves the

---

*for the Double Helix*, 26 AIPLA Q.J. 381, 418 (1998) ("making the inventions of *Amgen, Fiers* and *Lilly* today would be routine").

problem up one level of abstraction to argue that the Federal Circuit is not mistaken as a matter of fact about the state of technology in biotechnology and software, but rather mistaken as a matter of policy about the needs of those industries.  If the court is taking the trouble of fashioning individual patentability standards for different areas of subject matter, one would expect that the standards fashioned would be suited to the needs of the different areas addressed -- yet that seems not to be the case here.

Yet a third alternative explanation, at least for the biotechnology cases, might be found in a sort of judicial economy.  Rather than fashion new doctrines, the Federal Circuit biotechnology cases may be an attempt to extend existing chemical patent doctrines to encompass biotechnology.  Such an approach has the advantage of legal predictability, relying on established precedent with which innovators and their attorneys are already familiar.  But this explanation fails to explain why such doctrine should not also be extended to encompass software technology – why the explicit listing of computer code is not as critical to obviousness or disclosure as the explicit listing of genetic code seems to be.  Moreover, all these explanations beg the question of whether different technologies ought to have their own specific patenting standards, whether those standards are dictated by unique facts, or by legal considerations.

If, as we suggest, the concept of the PHOSITA makes sense, why has the Federal Circuit got it wrong in these industries?  We think there are several structural barriers that make it difficult for courts to accurately assess the level of skill in a complex technological art.  First, it is worth emphasizing that judges are at a rather serious disadvantage in trying to put themselves in the shoes of an ordinarily skilled scientist.  Judges ordinarily don't have any scientific background, and at least at the district court

*Is Patent Law Technology-Specific?* Burk & Lemley                                                                DRAFTlevel most law clerks don't either. Further, district court judges have extremely full dockets with many different types of cases. The average judge may hear no more than one patent case every few years.[114] Few of those will be software or biotechnology cases.[115] So a very busy judge must learn not only patent law but also some difficult science in a very short period of time. Expert witnesses can help, but the Federal Circuit has imposed some limits on the extent to which district courts can rely on such evidence.[116] In particular, courts must avoid the temptation to assume that the expert witness *is* a person ordinarily skilled in the art.[117] Even the Federal Circuit, which doesn't

---

[114] There are roughly 1700 patent cases filed per year. The exact data for the years 1995-1999 can be found in the Derwent Litalert database, available at http://www.derwent.com/intellectualproperty/litalert.html. The data that follow were compiled as of June 1, 2000, and involve cases labeled "patent."

| Year | Number of Cases Filed |
|---|---|
| 1999 | 1,652 |
| 1998 | 1,730 |
| 1997 | 1,731 |
| 1996 | 1,514 |
| 1995 | 1,258 |

Most of these cases settle, however. Kimberly Moore's comprehensive study of all patent cases that went to trial found only 1,411 cases in the 17 years from 1983 to 1999, an average of less than 100 cases per year. Kimberly A. Moore, *Judges, Juries, and Patent Cases – An Empirical Peek Inside the Black Box*, 99 **Mich. L. Rev.** 365, 380 (2000). Since there are over 600 district court judges in the United States, it is obvious that most judges get only a few filed patent cases a year, and well less than one patent trial a year. In fact, many judges get even fewer cases than this number would suggest (though others get more), since the concentration of innovation in certain regions and the permissibility of forum shopping in patent cases cause patent cases to be bunched in a few districts. *See* Kimberly A. Moore, *Forum Shopping in Patent Cases: Does Geographic Choice Affect Innovation?*, 79 **N.C. L. Rev.** 889 (2001) (analyzing where patent suits are filed).

[115] *See* John R. Allison & Mark A. Lemley, *Empirical Evidence on the Validity of Litigated Patents*, 26 **AIPLA Q.J.** 185, 217 & Table 5 (1998) (between 1989 and 1996, only 3% of patent cases involved biotechnology and only 1% involved software).

[116] *See* Vitronics Corp. v. Conceptronic, Inc., 90 F.3d 1576 (Fed. Cir. 1996) (courts may rely on expert testimony in construing patent claims only in rare circumstances); *but compare* Pitney-Bowes v. Hewlett-Packard Corp., 182 F.3d 1298, 1308 (Fed. Cir. 1999) (judges may hear expert testimony on the meaning of patent claims, but may not normally *rely* on such testimony). This distinction between admitting testimony to help the judge understand the claims and reliance on such testimony may make conceptual sense, but courts reading this line of cases may be reluctant to hear such evidence at all. Plus, it won't help much in deciding pretrial motions.

[117] *See, e.g.,* Dayco Prods. v. Total Containment, Inc., __ F.3d __ (Fed. Cir. July 20, 2001) ("our objective is to interpret the claims from the perspective of one of ordinary skill in the art, not from the viewpoint of counsel or expert witnesses . . ."); Endress + Hauser v. Hawk Measurement Sys., 122 F.3d 1040 (Fed. Cir.

suffer nearly so much from these limitations,[118] is not in a position to fully understand all of the science it encounters.[119] Given these limitations, courts understandably won't get it right all the time.[120]

Second, the timing of the PHOSITA analysis complicates the court's task. While the court will determine the level of skill in the art during a pretrial hearing or at trial, the appropriate level of skill in the art is not what people know at the time of trial, but what people knew at the time of the invention.[121] On average, it takes more than twelve years from the time a patent application is filed until final judgment on the merits; it takes even longer from the date of invention, of course.[122] So courts trying to determine the level of skill in the art must learn not just science, but the history of science. Courts and expert witnesses must shut out of their minds intervening developments in the field. This is

---

1997) ("The person of ordinary skill in the art is a theoretical construct . . . and is not descriptive of some particular individual"; experts need not themselves be of ordinary skill in the art).

[118] While relatively few Federal Circuit judges have technology backgrounds, see John R. Allison & Mark A. Lemley, *How Federal Circuit Judges Vote in Patent Validity Cases*, 27 **Fla. St. U. L. Rev.** 745, 751 n.23 (2000) (detailing the background of judges on the court in 1996), many of their clerks do. Further, the Federal Circuit has more time to consider each case, has the full record before it, and gets many more patent cases (and therefore more software and biotechnology cases) than any district court judge would.

[119] Arti Rai argues that the Federal Circuit should defer to the PTO, because the PTO better understands biotechnology. Rai, *supra* note __, at __. We agree with her that the Federal Circuit makes mistakes in this area. We are not persuaded that the PTO can do any better, however, particularly given the minimal time examiners can spend on any one invention. *See* Mark A. Lemley, *Rational Ignorance at the Patent Office*, 75 **Nw. U. L. Rev.** __ (forthcoming October 2001) (examiners spend 18 hours per application on average).

[120] *Cf.* Stephen L. Carter, *Custom, Adjudication, and Petrushevsky's Watch: Some Notes From the Intellectual Property Front*, 78 **Va. L. Rev.** 129, 132 (1992) (worrying that judges may not be particularly good at "judicial anthropology").

[121] *See* Arkie Lures, Inc. v. Gene Larew Tackle, Inc., 119 F.3d 953 (Fed. Cir. 1997) (PHOSITA analysis must "focus on conditions as they existed when the invention was made" in obviousness cases).

[122] Allison & Lemley, *Empirical Evidence*, *supra* note __, at 236 Table 11 (12.3 years on average). This has been a particular problem in biotechnology cases, particularly because they spend longer in prosecution. *See, e.g.,* Enzo Biochem v. Calgene, Inc., 188 F.3d 1362, 1371 (Fed. Cir. 1999) (16-year-old invention); Genentech, Inc. v. Novo Nordisk, 108 F.3d 1361, 1367 (Fed. Cir. 1997) (18-year-old

notoriously hard to do. Empirical evidence has demonstrated that people in general, and judges in particular, are subject to a "hindsight" bias: they are likely to reason backwards from what did happen to make assumptions about what was likely to happen *ex ante*.[123] The Federal Circuit has repeatedly recognized the problem of hindsight bias in its obviousness jurisprudence,[124] and has built rules designed to cope with it there,[125] but in fact hindsight bias risks infecting the PHOSITA analysis in enablement and claim scope as well. Hindsight bias will normally lead factfinders to overestimate the level of skill in the art, since subsequent advances will suggest that the invention couldn't have been that difficult to do. Occasionally, however, hindsight bias may have the opposite effect, notably where certain things known or believed at one time to be feasible turn out later to be more difficult than anticipated.[126]

---

invention); Jeffrey S. Dillen, *DNA Patentability – Anything But Obvious*, 1997 **Wisc. L. Rev.** 1023, 1038 (noting this time lag).

[123]  There is an interesting empirical literature in the behavioral law and economics movement on hindsight bias. The existence of such a bias is well documented. In the behavioral science literature, see, e.g., Baruch Fischhoff, *Hindsight & Foresight: The Effect of Outcome Knowledge on Judgment Under Uncertainty*, 1 **J. Experimental Psychol.: Hum. Perception & Performance** 288 (1975); Amos Tversky & Daniel Kahneman, *Availability: A Heuristic for Judging Frequency and Probability*, 5 **Cognitive Psych.** 207 (1973). In the legal literature, see, *e.g.,* **Cass Sunstein ed., Behavioral Law and Economics** (2000); Eric Talley, *Disclosure Norms*, 149 **U. Pa. L. Rev.** 1955, 2000 (2001); Russell Korobkin & Thomas S. Ulen, *Law and Behavioral Science: Removing the Rationality Assumption from Law and Economics*, 88 **Calif. L. Rev.** 1051, 1095 (2000). There is even empirical evidence that federal judges are subject to hindsight bias. *See* Chris Guthrie et al., *Inside the Judicial Mind*, 86 **Cornell L. Rev.** 777 (2001).

[124]  *See, e.g.,* Al-Site Corp. v. VSI Int'l, 174 F.3d 1308 (Fed. Cir. 1999); Monarch Knitting Machinery Corp. v. Fukuhara Indus. & Trading Co., 139 F.3d 1009 (Fed. Cir. 1998).

[125]  *See, e.g.,* In re Dembiczak, 175 F.3d 994 (Fed. Cir. 1999) ("Our case law makes clear that the best defense against the subtle but powerful attraction of a highsight-based obviousness analysis is rigorous application of the requirement for a showing of the teaching or motivation to combine prior art references."). Indeed, the Federal Circuit may have overcompensated, making it very difficult to combine references in order to prove obvious. *See* Lemley & O'Brien, *supra* note __, at 301 (making this argument). For an extremely strict statement of the legal standard on combining references, see Winner Int'l Royalty Corp. v. Wang, 202 F.3d 1340 (Fed. Cir. 2000).

[126]  For a detailed discussion of hindsight in biotechnology cases, see Lawrence M. Sung, *On Treating Past as Prologue*, 2001 **U. Ill. J. L., Tech. & Pol'y** 75.

Finally, the backward-looking nature of the legal system itself creates a problem that is in some sense the opposite of the hindsight bias. Legal rules are based on *stare decisis*. The law accumulates nuance over time by respecting and building on the body of existing precedent. Only rarely will courts expressly reject their prior decisions. This system has worked well over time in producing thoughtful legal rules.[127] Judges trained in this process will naturally tend to apply it to factual issues they see repeatedly as well. Indeed, doing so seems economical as well, since revisiting those factual determinations appears redundant. Thus, once the Federal Circuit has ruled on the level of skill in a particular art, the temptation is strong for both that court and district courts to apply that determination in subsequent cases as well. This tendency is evident in both software and biotechnology cases. In the software cases, the court in *Northern Telecom* held that the patentee needn't disclose the actual code implementing a program in order to satisfy the enablement or best mode requirements. The court in that case acknowledged that determinations of the level of skill in the computer industry should be made on a case-by-case basis.[128] But subsequent Federal Circuit decisions have not inquired separately into the level of skill in the art, or explored the complexity of the program before them in much detail. Instead, they have tended to rely on prior cases holding that code need not be disclosed. In biotechnology the linkage is even stronger. *In re Bell* concluded that knowledge of an amino acid sequence produced by a gene, coupled with a plan for identifying the DNA sequence of the gene, did not render the DNA sequence itself obvious.[129]  *In re Deuel* relied on *Bell*'s conclusion, despite the fact that biotechnology had advanced somewhat between the two inventions.[130]  And in *Regents of the University of California v. Eli Lilly & Co.*,[131] the court expressly relied on its conclusions about the level of skill in the art in *Bell* and *Deuel* to determine its conclusions regarding written description.[132]  *Fiers* is even more explicit in this regard, creating a firm rule that conception of a DNA sequence requires a listing of that sequence "regardless of the complexity or simplicity of the method of isolation."[133]

---

[127]   For arguments suggesting the common law evolves towards efficiency over time, see **Richard A. Posner, Economic Analysis of Law** 23-27 (1st ed. 1979); George L. Priest, *The Common Law Process and the Selection of Efficient Rules*, 6 **J. Legal Stud.** 65 (1977); Paul H. Rubin, *Why is the Common Law Efficient?*, 6 **J. Legal Stud.** 51 (1977).

[128]   908 F.2d 931, 941 (Fed. Cir.), cert. denied, 111 S. Ct. 296 (1990).

While apparently logical, the reliance on industry-specific precedent in determining the level of skill in the art is problematic.  First, while both obviousness and enablement rely on the PHOSITA construct, the PHOSITA is not necessarily the same for obviousness and enablement even in a single case.  Obviousness is tested at the time the invention was made, while enablement is tested at the time the application was filed. Clearly the application cannot be filed until after the date of invention, and in some cases several years elapse between the two.[134] The knowledge in the art can change during this period, sometimes dramatically. Second, and more important, the level of skill in the art will normally change between the dates of different inventions. It is hazardous, therefore, to rely on one court's statement of the level of skill in the art as determinative or even evidentiary of the level of skill in the same art at a different time. The level of skill in the art is a factual question that must be determined anew on the particulars of each case.[135]

---

[129]   In re Bell, 991 F.2d 781 (Fed. Cir. 1993).

[130]   In re Deuel, 51 F.3d 1552, 1559 (Fed. Cir. 1995).  In *Bell*, the prior art disclosed the amino acid sequence for the proteins of interest, and a method for cloning genes.  By contrast, the art in *Deuel* disclosed only a partial sequence.  Nontheless, the passage of time between the priority dates of the applications – almost 10 years [check] – was ignored by the court, which did not focus on or even mention when the inventions occurred.

[131]   Regents of the University of California v. Eli Lilly & Co., 119 F.3d 1559 (Fed. Cir. 1997).

[132]   *Id*. at 1567:
> Example 6 provides the amino acid sequence of the human insulin A and B chains, but that disclosure also fails to describe the cDNA. Recently, we held that a description which renders obvious a claimed invention is not sufficient to satisfy the written description requirement of that invention. Lockwood, 107 F.3d at 1572, 41 USPQ2d at 1966. We had previously held that a claim to a specific DNA is not made obvious by mere knowledge of a desired protein sequence and methods for generating the DNA that encodes that protein. See, e.g., In re Deuel, 51 F.3d 1552, 1558, 34 USPQ2d 1210, 1215 (1995) ("A prior art disclosure of the amino acid sequence of a protein does not necessarily render particular DNA molecules encoding the protein obvious because the redundancy of the genetic code permits one to hypothesize an enormous number of DNA sequences coding for the protein."); In re Bell, 991 F.2d 781, 785, 26 USPQ2d 1529, 1532 (Fed.Cir.1993). Thus, a fortiori, a description that does not render a claimed invention obvious does not sufficiently describe that invention for purposes of § 112, ¶ 1. Because the '525 specification provides only a general method of producing human insulin cDNA and a description of the human insulin A and B chain amino acid sequences that cDNA encodes, it does not provide a written description of human insulin cDNA.

[133]   Fiers v. Revel, 984 F.2d 1164, 1169 (Fed. Cir. 1993).

### III.     Should Patent Law Be Technology-Specific?

If we are right that the nominally unified standards of patent law are in fact technology-specific because they rely so heavily on the level of skill in the art, the question remains whether this heterogeneity is desirable. In the last section, we suggested that the Federal Circuit has, either as a matter of fact or as a conclusion of law, improperly characterized the level of skill in the art in the biotechnology and software industries. But even if these problems are overcome, there seems little question that the effect of reliance on the PHOSITA will be to impose different standards in different industries. In section A, we discuss the policy implications of this approach for software and biotechnology. In section B, we discuss the more general implications of patent-specificity for patent reform.

---

[134]   The law permits a one year grace period between any public act and the filing of a patent application. 35 U.S.C. § 102(b). But many inventors wait even longer between invention and the filing of an application. This is permissible, so long as they do not put the invention on sale or in public use in the interim, and do not abandon it. 35 U.S.C. § 102(c).

[135]   For a detailed discussion, see Dillen, *supra* note __, at 1039-44. The CCPA recognized this in *In re Driscoll*, 562 F.2d 1245, 1250 (C.C.P.A. 1977), and the Federal Circuit in Enzo Biochem v. Calgene, Inc., 188 F.3d 1362, 1374 n.10 (Fed. Cir. 1999). But it has proven a hard rule to adhere to.

### A.     The Proper Scope of Software and Biotechnology Patents

Nonobviousness is a function of uncertainty.[136] Where uncertainty is higher, courts should lower the standard of patentability to compensate for the risk of failure (and therefore the lower expected reward per dollar invested). While courts have traditionally focused on uncertainty and hence obviousness as a function of invention, in fact invention is rewarded in the marketplace only to the extent it is embodied in a successful commercial product that can be sold at a price above marginal cost. Getting from an invention to a successful product requires many more steps: developing the product, testing it, producing it, marketing it, and in many cases developing complementary products or even whole new industries that can take advantage of the invention in the most efficient way. The entire process of research, development, and turning an idea into a finished product can be described as innovation. Invention is thus a subset of innovation.[137]

Under this model of obviousness, uncertain and high-cost innovation – not just invention – should more likely be entitled to a determination of nonobviousness. High cost will tend to correlate with higher risks, as the larger investment increases the opportunity for loss at any probability of success. The greater variance in outcomes might be expected to deter the rational entrepreneur from investing in such high cost projects unless the expected reward is correspondingly greater. This higher level of perceived risk may be to some extent counterbalanced by lowering the standard for patentability of high-cost projects, increasing the likelihood of financial reward.

---

[136] For a detailed elucidation of the ideas in this paragraph, see Robert P. Merges, *Uncertainty and the Standard of Patentability*, 7 **High Tech. L.J.** 1 (1992).

Application of this innovation-based model to software and biotechnology reveals problems with the current legal standards. As we have seen, the courts have portrayed computer programming as relatively straightforward and biotechnology as more uncertain. As a result, software patents are more likely to be found obvious, but if valid tend to be broad in scope and require only minimal disclosure.[138] Biotechnology patents, by contrast, are relatively easy to prove nonobvious but require detailed disclosure and are extremely narrow.[139] Arguably this result is exactly backwards as a matter of innovation policy. As we shall see, common policy arguments responding to market conditions peculiar to the two industries call for narrow software patents and broad biotech patents. At a minimum, the PHOSITA test -- even if correctly applied -- will not produce the results that seem optimal for each industry.

### 1.      Optimal Software Patent Policy

The computer industry is characterized by a large number of rapid, iterative improvements on existing products.[140] Computer programs normally build on preexisting

---

[137]   In using this typology, we follow William Kingston. *See* **William Kingston ed., Direct Protection of Innovation** (1987).

[138]   *See supra* notes __-__ and accompanying text.

[139]   *See supra* notes __-__ and accompanying text.

[140]   *See, e.g.,* Cohen & Lemley, *supra* note __, at 40-42; Menell, *supra* note __; Samuelson et al., *supra* note __.

*Is Patent Law Technology-Specific?* Burk & Lemley     DRAFT*Is Patent Law Technology-Specific?* Burk & Lemley                                                                                       DRAFT ideas, and often on prior code itself.[141] This incremental improvement is desirable for a variety of reasons. First, it responds to the hardware-based architectural constraints of the software industry. Data storage capacity, processing speed, and transmission rates have all increased steadily over time.[142] Programs written during an older period therefore faced capacity constraints that disappear over time. It makes sense to improve those products progressively as the constraints that limit the functionality of the programs disappear. Second, incremental improvement of existing programs and ideas tends to render programs more stable. It is received wisdom that you should avoid version 1.0 of any software product, because its maker is unlikely to have all the bugs worked out. Iterative programs built on a single base tend to solve these problems over time. This is most obviously true when actual computer code is reused, but it is true even when tested algorithms or structures are replicated in new programs. Third, iterative improvement helps preserve interoperability, both among generations of the same program and across programs.[143]

---

[141] On reuse of existing code, both within and across companies, see Lemley & O'Brien, *supra* note __.

[142] Moore's "law" provides that historically the speed of microprocessors has doubled every 18 months. It is well known that data storage capacity and transmission rates have shown similarly exponential increases.

[143] For the same reason, reverse engineering has had a respected place as a legitimate means of creating interoperability. Virtually all recent copyright decisions have endorsed reverse engineering in some circumstances. *E.g.*, DSC Communications Corp. v. DGI Techs., Inc., 81 F.3d 597, 601 (5th Cir. 1996); Bateman v. Mnemonics, Inc., 79 F.3d 1532, 1539 n.18 (11th Cir. 1996); Lotus Dev. Corp. v. Borland Int'l, Inc. 49 F.3d 807, 817-18 (1st Cir. 1995) (Boudin, J., concurring); Atari Games Corp. v. Nintendo of America, Inc., 975 F.2d 832, 843-44 (Fed. Cir. 1992); Sega Enters. Ltd. v. Accolade, Inc., 977 F.2d 1510, 1527-28 (9th Cir. 1992); Vault Corp. v. Quaid Software Ltd., 847 F.2d 255, 270 (5th Cir. 1988); DSC Communications Corp. v. Pulse Communications, Inc., 976 F. Supp. 359 (E.D. Va. 1997); Mitel, Inc. v. Iqtel, Inc., 896 F. Supp. 1050 (D. Colo. 1995), *aff'd on other* grounds, 124 F.3d 1366 (10th Cir. 1997). On the other hand, a few early decisions rejected compatibility as a justification for copying. *E.g.,* Apple Computer, Inc. v. Franklin Computer Corp. 714 F.2d 1240 (3d Cir. 1983); Digital Communications Ass'n v. Softklone Distrib. Corp., 659 F. Supp. 449 (N.D. Ga. 1987); *cf.* DSC Communications Corp. v. Pulse Communications, Inc., 170 F.3d 1354 (Fed. Cir. 1999) (acknowledging the right to reverse engineer for some purposes, but holding it unjustified in this case).

    As with courts, the overwhelming majority of commentators endorse a right to reverse engineer copyrighted software, at least for certain purposes. *E.g.*, **Jonathan Band & Masanobu Katoh,**

The software industry also has relatively low fixed costs and a short time to market. The archetypal software invention is one made by two people working in a garage.[144] While the costs of writing software have increased substantially over time as programs became more complex, the costs of writing and manufacturing computer programs are still low relative to the fixed costs of development in many industries. Further, computer program life cycles are rapid. Unlike industries like steel or aircraft, where new generations of products are infrequent and products may last for decades, computer programs tend to be replaced every few years, often by new versions of the same program.

The implications of these economic characteristics for patent law are threefold. First, the need for strong patent protection is somewhat less for software inventions than

---

**Interfaces on Trial: Intellectual Property and Interoperability in the Global Software Industry** 167-226 (1995); Cohen, *supra* note __; Lawrence D. Graham & Richard O. Zerbe, Jr., *Economically Efficient Treatment of Computer Software: Reverse Engineering, Protection, and Disclosure*, 22 **Rutgers Computer & Tech. L.J.** 61 (1996); Dennis S. Karjala, *Copyright Protection of Computer Software, Reverse Engineering, and Professor Miller*, 19 **U. Dayton L. Rev.** 975, 1016-18 (1994); Maureen A. O'Rourke, *Drawing the Boundary Between Copyright and Contract: Copyright Preemption of Software License Terms*, 45 **Duke L.J.** 479, 534 (1995); David A. Rice, *Sega and Beyond: A Beacon for Fair Use Analysis . . . At Least as Far As It Goes*, 19 **U. Dayton L. Rev.** 1131, 1168 (1994); Pamela Samuelson, *Fair Use for Computer Programs and Other Copyrightable Works in Digital Form: The Implications of Sony, Galoob and Sega*, 1 **J. Intell. Prop. L.** 49 (1993); Tyler G. Newby, Note, *What's Fair Here Is Not Fair Everywhere: Does the American Fair Use Doctrine Violate International Copyright Law?*, 51 **Stan. L. Rev.** 1633, 1657-58 (1999); Timothy Teter, Note, *Merger and the Machines: An Analysis of the Pro-Compatibility Trend in Computer Software Copyright Cases*, 45 **Stan. L. Rev.** 1061 (1993) (arguing that the value of computer programs depends on interoperability); *see also* Pamela Samuelson & Suzanne Scotchmer, *The Law and Economics of Reverse Engineering,* (working paper 2001) (suggesting that reverse engineering should be legal when it promotes interoperability, but not when it permits free riding); Cohen & Lemley, *supra* note __, at 17-21 (expressing concern that patent law may not protect reverse engineering).

For a contrary view, see generally Anthony L. Clapes, *Confessions of an Amicus Curiae: Technophobia, Law and Creativity in the Digital Arts*, 19 **U. Dayton L. Rev.** 903 (1994) (contending that there should be no right to reverse engineer software), and Arthur R. Miller, *Copyright Protection for Computer Programs, Databases, and Computer-Generated Works: Is Anything New Since CONTU?*, 106 **Harv. L. Rev.** 977 (1993) (same).

[144] Hewlett and Packard and Jobs and Wozniack are the classic examples, but the story has taken on a life of its own. *See, e.g.,* Micalyn S. Harris, *UCITA: Helping David Face Goliath*, 18 **J. Marshall J. Comp. & Info. L.** 365, 375 (1999).

it is in other industries. Software patents are important, but the relatively low fixed costs associated with software development, coupled with other forms of overlapping intellectual property protection for software,[145] mean that innovation in software does not depend critically on strong, broad protection. Second, the rapid, incremental innovation crucial to the software industry may be stifled or at least regulated by older companies that own software patents based on prior generations of products. Cohen and Lemley offer several reasons to believe that the doctrine of equivalents may be applied too broadly in the software industry, allowing owners of old software patents to prevent the development of new generations of technology.[146] Finally, a culture of rapid-fire incremental improvements leads to a large number of low-level innovations. Copyright is not capable of providing effective protection for such innovations because it does not protect functionality.[147] Some form of protection for such innovations is desirable. In the

---

[145] Predominantly copyright, but also trade secret and contract law. One factor militating in favor of stronger intellectual property protection in software is the ease of duplication of digital information in the networked world. But copyright protection is much better suited than patent to preventing exact duplication. Copyright law has also been modified to better prevent such copying in the computer context by allowing copyright owners to control access to copy-protected works. *See* 17 U.S.C. §1201 (the Digital Millenium Copyright Act).

[146] Cohen & Lemley, *supra* note __, at 39-50 (incremental nature of software innovation, lack of good prior art, rapid pace of change, and the difficulty of characterizing code inventions in words all contribute to broad readings of software inventions). They write:
> The pattern of cumulative, sequential innovation and reuse that prevails in the software industry creates the risk that software patents will cast large shadows in infringement litigation. Specifically, we believe that because innovation is especially likely to proceed by building on existing code in other programs, the temptation for the trier of fact to find equivalence of improvements will be correspondingly greater.

*Id*. at 41.

[147] For a detailed discussion, see Samuelson et al., *supra* note __, at 2350-56; Pamela Samuelson, *CONTU Revisited: The Case Against Copyright Protection for Computer Programs in Machine-Readable Form*, 1984 **Duke L.J.** 663.

absence of other forms of protection, a large number of narrow software patents may be the best way of protecting these low-level innovations.[148]

Patent protection for such incremental software inventions should be relatively narrow, and in particular should not generally extend across several product generations.[149] But broad software patents are precisely what the court's PHOSITA approach has produced. By defining software inventions in broad terms, supported by very little in the way of detailed disclosure, the Federal Circuit has encouraged software patents to be drafted broadly and to be applied to accused devices that are far removed from the original patented invention. By implication, the Federal Circuit's standard also seems to suggest that many software patents on low-level incremental improvements will be invalid for obviousness. In software, then, the Federal Circuit's standard seems to have it backwards.

### 2.     Optimal Biotechnology Patent Policy

If any technology fits the criteria of high cost and high risk innovation, it is certainly biotechnology. Biotechnology products appear in a wide variety of economic sectors, from pharmaceuticals to foodstuffs to industrial processes.[150] Development of biotechnology products, particularly in the pharmaceutical sector, has been characterized by extremely long development times and high development costs. Such delays are due in part due to the stringent regulatory oversight exercised over the safety of new drugs, foods, biologics, and over environmental release of new organisms.[151] Yet the onerous regulatory requirements to which biotechnology is subject may obscure a more fundamental uncertainty that justifies such oversight: biotechnology products arise out of living systems, and are typically intended to interact with other human or non-human living systems. Such interactions, whether physiological or ecological, are enormously complex and the systems involved poorly characterized. As a consequence, the functionality of biotechnology products is always unforeseeable, and always involves a high degree of uncertainty and risk.[152]

---

[148]   The Manifesto worries that software patents may be too broad given the incremental nature of software innovation. Samuelson et al., *supra* note __, at 2345-46. *See also* Pamela Samuelson, *Benson Revisited: The Case Against Patent Protection for Algorithms and Other Computer Program-Related Inventions*, 39 **Emory L.J.** 1025 (1990). As noted below, we share this concern, but believe the solution is to narrow the scope of those patents.

[149]   *See generally* Richard R. Nelson, *Intellectual Property Protection for Cumulative Systems Technology*, 94 **Colum. L. Rev.** 2674 (1994).

[150]   *See* Dan L. Burk, *A Biotechnology Primer*, 55 **U. Pitt. L. Rev.** 611 (1994).

Consistent with these characteristics, the current Federal Circuit jurisprudence lowers the obviousness barrier for biotechnology. Much of the criticism that has been directed against the biotechnology cases described above has focused on the availability of research tools that have made routine the isolation and characterization of biological macromolecules.[153] Given such tools, the outcome of a search for a particular nucleotide or protein seems relatively certain, and hence it is argued, obvious. But to the extent that patents spur innovation, rather than invention, a relatively low threshold for patentability is still needed for biotechnology.[154] The availability or unavailability of a patent is expected to have little effect on the incentive to engage in preliminary research[155] – that is, in the case of biotechnology, to use the available tools to secure a macromolecule of interest. But the ready availability of tools for finding a new biotechnology product do not change the high cost and uncertainty entailed in developing a marketable product using that macromolecule; hence a lowered standard of obviousness still makes sense from a policy standpoint.[156]

---

[151]  PharmA estimates that the total time spent from the beginning of a research project to the marketing of a successful drug is 14.2 years, 1.8 years of which is due to the FDA approval process. *See* http://www.phrma.org/publications/publications/profile01/chapter2.pdf. Estimates of the average cost of drug development and testing range from $110 million to $500 million; the latter is the industry's figure. *Compare id*. with http://www.citizen.org/Press/pr-drugs33.htm.

[152] For example, the Centocor sepsis antibody, a highly promising biotechnology treatment, succeeded in passing many years of costly trials,but failed in the last phase of FDA approval.

[153] Jonathan M. Barnett, *Cultivating the Genetic Commons: Imperfect Patent Protection and the Network Model of Innovation,* 37 **San Diego L. Rev.** 987 (2000); Philippe Ducor, *New Drug Discovery Technologies and Patents*, 22 **Rutgers Comp. & Tech. L.J.** 369 (1996); Arti K. Rai, *Intellectual Property Rights in Biotechnology: Addressing New Technology*, 34 **Wake Forest L. Rev.** 827 (1999). *See generally* John M. Golden, *Biotechnology, Technology Policy, and Patentability: Natural Products and Invention in the American System*, 50 **Emory L.J.** 101 (2001).

Yet what the Federal Circuit gives biotechnology with one hand, it takes away with the other. Although biotechnology patents are relatively easy to obtain under the obviousness standard, the accompanying enablement and written description standards dramatically narrow the scope of the resulting patents. By requiring disclosure of the particular structure or sequence in order to claim biological macromolecules, the Federal Circuit effectively limits the scope of a patent on those molecules to the structure or sequence disclosed. This standard concomitantly dictates that the inventor have the molecule "in hand" (so to speak) before being able to claim it – in other words, after a substantial investment has already been made in isolating and characterizing the molecule. The result is that everyone who invests in discovering a new molecule will receive a patent, but one that is trivial to evade. Under this standard, no one is likely to receive a patent broad enough to support the further costs of development.[157]

---

[154] *See* Merges, *Uncertainty*, *supra* note __; Robert P. Merges, *One Hundred Years of Solicitude: Intellectual Property Law, 1900-2000*, 88 **Calif. L. Rev.** 2187, 2225-27 (2000); Karen I. Boyd, *Nonobviousness and the Biotechnology Industry: A Proposal for a Doctrine of Economic Nonobviousness*, 12 **Berkeley Tech. L.J.** 311 (1997).

[155] *See, e.g.,* **Robert P. Merges, Patent Law and Policy** 519 (2d ed. 1997).

[156] One way to think of this is to reconceive patents as a financing mechanism: by providing definable rights, patents enable companies to obtain the funding they need to turn an invention into a product. *See* Golden, *supra* note __, at 167-172; Mark A. Lemley, *Reconceiving Patents in the Age of Venture Capital*, 4 **J. Sm. & Emerging Bus. L.** 137 (2000).

An additional drawback to the proliferation of narrow biotechnology patents may be the development of what has been termed an "anti-commons" in the biotechnology industry.[158] The anticommons is characterized by fragmented property rights, the aggregation of which is necessary to make effective use of the property.[159] Aggregating such fragmented property rights entails high search and negotiation costs to locate and bargain with the many rights owners whose collective permissions are necessary to complete broader development. This type of licensing environment may quickly become dominated by "hold outs" who refuse to license their essential sliver of the pie unless bribed.[160] Because a given project will fail without their cooperation, "hold-outs" may be prompted to demand a bribe close to the value of the entire project.[161] And, of course,

---

[157] *See* Kenneth G. Chahine, *Enabling DNA and Protein Composition Claims: Why Claiming Biological Equivalents Encourages Innovation*, 25 **AIPLA Q.J.** 333 (1997) (arguing for a broader scope of biotechnology patents, extending to proteins with comparable biological activity).
Curiously, Merges doesn't see this as a major problem, suggesting that in general "the Federal Circuit has overall been quite successful at integrating biotechnology cases into the fabric of patent law." Merges, *Solicitude*, *supra* note __, at 2228. We think the written description cases and the correspondingly narrow scope afforded biotechnology patents are a more serious problem than Merges acknowledges.

[158] Michael A. Heller & Rebecca S. Eisenberg, *Can Patents Deter Innovation? The Anticommons in Biomedical Research*, 280 **Sci.** 698, 698 (1998) (patenting genetic research can lead to an "anticommons" in which multiple, conflicting property rights impede efficient use of the patents). *see also* Michael A. Heller, *The Tragedy of the Anticommons: Property in the Transition From Marx to Markets*, 111 **Harv. L. Rev.** 621 (1998); Rebecca S. Eisenberg, *A Technology Policy Perspective on the NIH Gene Patenting Controversy*, 55 **U. Pitt. L. Rev.** 633 (1994). *But see* John J. Doll, *The Patenting of DNA*, 280 **Sci.** 689 (1998) (noting that similar concerns have been raised with other technologies, such as polymers).
On the antitrust issues raised by patents on research tools, see John H. Barton, *Patents and Antitrust: A Rethinking in Light of Patent Breadth and Sequential Innovation*, 65 **Antitrust L.J.** 449 (1997). For discussion of a related anticommons problem, reach-through royalty agreements, see James Gregory Cullem, *Panning for Biotechnology Gold: Reach-Through Royalty Damage Awards for Infringing Uses of Patented Molecular Sieves*, 39 **Idea** 553 (1999).

[159] Heller, *supra* note __, at 670-72.

[160] On the holdout problem, see generally **Mancur Olson, The Logic of Collective Action** (1961).

[161] Lloyd Cohen, *Holdouts and Free Riders*, 20 J. Legal Stud. 351 (1991).

every property holder needed for the project is subject to this same incentive; if everyone holds out, the cost of the project will rise substantially, and probably will rise prohibitively

Unfortunately, if patent law is to accommodate the uncertainty of biotechnology innovation, this proliferation of narrow biotechnology patents may be nearly impossible to avoid under the reciprocal structure of obviousness and enablement in the PHOSITA patent doctrine.[162] In order for the invention to avoid obviousness, it must be deemed beyond the skill of the PHOSITA to construct given the level of disclosure in the prior art. Yet this means that in disclosing the invention, the inventor must tell those of ordinary skill a good deal more about how to make and use it, effectively raising the standard for enablement and written description. The Federal Circuit's insistence that the results biotechnology research are unforeseeable or unpredictable avoids the problem of obviousness, but this inevitably results in an extremely stringent standard for disclosure and description.

### B.      Can Patent Law Produce Optimal Patent Policy?

If we are right that patent law doctrines lead to troubling policy results in both software and biotechnology, what is to be done? In particular, the question is whether the law should adapt to our policy prescriptions, and if so how. In this section, we consider several possible ways of modify patent law to accommodate the dictates of patent policy.

---

[162] *See, e.g.,* Mark J. Stewart, *The Written Description Requirement of 35 U.S.C. § 112(1): The Standard After Regents of the University of California v. Eli Lilly & Co.*, 32 **Ind. L. Rev.** 537, 557-58 (1999) (noting

1.      **Industry-Specific Patent Legislation**

One obvious response to the different policy prescriptions described above is to explicitly legislate different patent standards for different industries. While patent law has historically been uniform, with a single set of legal standards designed to cover "everything under the sun made by man,"[163] Congress has shown increased interest in tailoring patent law to the needs of particular industries. In the last twenty years, it has lengthened the patent term for most pharmaceutical patents,[164] protected certain experimental uses of pharmaceuticals by generic suppliers from liability,[165] prohibited enforcing patents on medical procedures against doctors,[166] relaxed the obviousness standard for biotechnological processes,[167] and created a new defense against business method patents.[168] It has supplemented patent protection for semiconductors with a *sui generis* statute.[169] It has even passed a "private" patent bill lengthening the term of one narrow group of patents.[170] In each case, Congress reacted to particular complaints about the perceived unfairness of applying a general legal standard to a particular industry.

---

the linkage between the Federal Circuit's view of biotechnology as an uncertain art and the narrowness of the patents that result).

[163]   Diamond v. Chakrabarty, 447 U.S. 303, 309 (1980). *See also* TRIPs art. 27(1) (requiring that patents be available without discrimination as to the form of technology).

[164]   *See* 35 U.S.C. §§ 155, 156.

[165]   35 U.S.C. § 271(e).

[166]   *See* 35 U.S.C. § 287.

[167]   35 U.S.C. § 103(b).

[168]   35 U.S.C. § 271(a)(3).

[169]   Semiconductor Chip Protection Act, 17 U.S.C. §§ 901 et seq.

[170]   *See* 35 U.S.C. § 155A. On the history of private patent legislation, see Robert P. Merges & Glenn Harlan Reynolds, *The Proper Scope of the Copyright and Patent Power*, 37 **Harv. J. Legis.** 45 (2000).

Still other bills currently pending in Congress would change the patent law standards for business method patents or extend the patent for Claritin.[171]

A number of scholars have suggested that patent law needs to be modified to take account of the particular needs of the software industry. Some suggest that software patents are inappropriate altogether,[172] others that Internet business method patents are.[173] Others suggest that a form of *sui generis* patent-like protection is appropriate.[174] Still others who endorse the general framework argue that the courts should apply patent law to software in somewhat different ways than they do in other industries.[175] Similarly, scholars have suggested that biotechnology patent standards should deviate from the general patent law rules.[176] Some argue that certain types of biotechnological patents should be unpatentable altogether.[177] Others suggest that the disclosure requirements

---

[171] Cite Boucher-Berman; Claritin patent relief.

[172] *See, e.g.,* Samuelson, *Benson Revisited*, *supra* note __.

[173] *See, e.g.,* Matthew G. Wells, *Internet Business Method Patent Policy*, 87 **Va. L. Rev.** 729, 770-72 (2001).

[174] *See, e.g.,* Samuelson et al, *supra* note __; Peter S. Menell, *Tailoring Legal Protection for Computer Software*, 39 **Stan. L. Rev.** 1329 (1987). For a more recent and somewhat different proposal, see Lester Thurow, *Needed: A New System of Intellectual Property Rights*, **Harv. Bus. Rev.** 95, 95 (Sept.-Oct. 1995) (discussing software and biotechnology industries).

[175] Most commonly, people suggest that the rapid market cycles in software justify shorter terms of protection for software patents. Cites. *Cf.* Cohen & Lemley, *supra* note __, (suggesting ways to avoid overbroad application of the doctrine of equivalents and protect reverse engineering); Stern, *supra* note __; Cohen, *supra* note __ (both suggesting application of an innovative programmer standard to software patents).

[176] For a critical analysis of such proposals, see Dan L. Burk, *Biotechnology and Patent Law: Fitting Innovation to the Procrustean Bed*, 17 **Rutgers Comp. & Tech. L.J.** 1 (1991). As noted above, the rules in biotechnology cases already do diverge from the general rules in the context of obviousness standards for biotechnological processes. *See supra* note __.

[177] *See, e.g.,* Kojo Yelpaala, *Owning the Secret of Life: Biotechnology and Property Rights Revisited*, 32 **McGeorge L. Rev.** 111 (2000); Mark O. Hatfield, *From Microbe to Man*, 1 **Animal L.** 5 (1995) (both making moral arguments against patenting life). For a very different argument against the patenting of cDNA sequences, see Rebecca S. Eisenberg & Robert P. Merges, *Opinion Letter as to the Patentability of*

should be loosened,[178] that the obviousness standard should be lowered,[179] or that the scope of DNA sequence patents should be restricted.[180] They have variously argued that the Federal Circuit should defer to the PTO,[181] or conversely that the PTO should defer to the Federal Circuit.[182]

Calls to modify patent law are a natural response to the different effects patent law has in different industries. The economic effects of patents are quite different in different industries. Thus, in a perfect world the patent system might well be tailored to give optimal incentives to each different industry.

We do not live in a perfect world, however. In the real world, a number of factors caution against tailoring the patent law to the needs of particular industries. The most obvious barrier is legal – the TRIPs agreement prohibits member states from discriminating in the grant of patents based on the type of technology at issue.[183] As

---

*Certain Inventions Associated With the Identification of Partial cDNA Sequences*, 23 **AIPLA Q.J.** 1 (1995).

[178] *See, e.g.,* Hodges, *supra* note __; Janice M. Mueller, *The Evolving Application of the Written Description Requirement to Biotechnological Inventions*, 13 **Berkeley Tech. L.J.** 615 (1998); Harris A. Pitlick, *The Mutation on the Description Requirement Gene*, 78 **J. Pat. & Trademark Ofc. Soc'y** 209 (1998); Margaret Sampson, *The Evolution of the Enablement and Written Description Requirements Under 35 U.S.C. § 112 in the Area of Biotechnology*, 15 **Berkeley Tech. L.J.** 1233 (2000); Cliff D. Weston, *Chilling of the Corn: Agricultural Biotechnology in the Face of U.S. Patent Law and the Cartagena Protocol*, 4 **J. Sm. & Emerging Bus. L.** 377, 389-92 (2000).

For a related argument, that the biotech written description cases are really about enablement and serve to obscure the real purposes of the written description requirement, see Mark D. Janis, *On Courts Herding Cats: Contending With the "Written Description" Requirement (and Other Unruly Patent Disclosure Doctrines)*, 2 **Wash. U. J. L. & Pol'y** 55 (2000).

[179] *See* Boyd, *supra* note __, at 311.

[180] *See, e.g.,* Heller & Eisenberg, *supra* note __; Eisenberg & Merges, *supra* note __.

[181] *See* Rai, *supra* note __, at 842-47.

[182] *See* Craig R. Miles, *Goldilocks Patent Protection for DNA Inventions: Not Too Thick, Not Too Thin, But Just Right*, 2 **Modern Trends in Intell. Prop.** 3 (1998).

[183] GATT TRIPs, art. 27(1).

noted above, however, the United States has not faithfully followed this treaty mandate. Neither has the EU, which has industry-specific rules for compulsory licensing of pharmaceuticals and for the patentability of software and business methods.[184]

Even if industry-specific patent legislation were legal, we are not fully persuaded that it is a good idea. First, while economics can make useful policy suggestions as to how patents work in different industries, we are skeptical of its ability to dictate in detail the right patent rules for each industry.[185] Economic theory is more useful in making general suggestions about how the patent system can be adapted than it is as the basis for a whole series of new statutes.

Second, rewriting the patent law for each industry would involve substantial administrative costs and uncertainties. Congress would have to write new statutes not just for biotechnology and software, but for any number of different industries with special characteristics: semiconductors, chemistry, Internet, perhaps telecommunications and other industries would need separate statutes. District court judges, who already have enough trouble learning the arcane rules of patent law in the relatively few patent cases they hear, would have to learn a host of new statutes. The law supporting these statutes would be slow to develop, since fewer cases would come up involving each statute. The resulting uncertainty would perhaps enrich lawyers, but it surely would not be conducive to encouraging innovation. There will also be a great deal of line-drawing to be done, as the boundaries between industries are not clear-cut and are notoriously mutable. Semiconductor manufacturers patent and use software all the time. Drug delivery systems might be thought of as medical devices, pharmaceuticals, or biotechnology; presumably a different law would apply depending on how the invention was characterized. Even technologies that seem radically different, like biotechnology and software, may unexpectedly converge, as recent developments in bioinformatics and proteomics have made clear.[186] Further, a significant percentage of inventions fall into more than one field.[187] And of course new fields arise regularly; imagine trying to fit all modern inventions into categories created 50 or 100 years ago. As a result, it will prove impossible to carve up innovation into static fields.

---

[184] cites

[185] *See* Louis Kaplow, *The Patent-Antitrust Intersection: A Reappraisal*, 97 **Harv. L. Rev.** 1813 (1984) (rejecting as impractical an effort to determine optimal patent length for each industry).

This point raises a related one. The history of industry-specific statutes suggests that many turn out to be failures because they are drafted with current technology in mind, and are not sufficiently general to accommodate the inevitable change in technology. The best example is the Semiconductor Chip Protection Act.[188] Passed after six years of debate, the SCPA created a detailed set of rules designed to protect semiconductor mask works. But it has virtually never been used.[189] The most likely reason is that the particular focus of the SCPA -- duplication of mask works -- is obsolete because of changes in the way semiconductor chips are made. Industry-specific patent statutes risk a similar fate.[190]

---

[186] Bioinformatics involves the regularized use of computer models to identify and predict gene structures. *See, e.g.,* Ken Howard, *The Bioinformatics Gold Rush*, **Sci. Am.,** July 2000, at 58. Proteomics involves the use of computer chips to build and test proteins. *See, e.g.,* Carol Ezzell, *Beyond the Human Genome*, **Sci. Am.,** July 2000, at 64, 67-69 (describing proteomics).

[187] *See* Allison & Lemley, *Who's Patenting What*, *supra* note __, at 2114 n.45 (on average, patents in the late 1990s fell into 1.49 different technology areas). This is actually a modest increase from the 1970s, when the number was 1.37. Most of this increase is due to the growth of software and biotechnology patents. *See* Allison & Lemley, *Complexity*, *supra* note __, at [draft at 20 Table 1].

[188] 17 U.S.C. §§901 et seq.

[189] There is only one reported case interpreting the SCPA. Brooktree Corp. v. Advanced Micro Devices, 977 F.2d 1555 (Fed. Cir. 1992).

Finally, and of most concern, technology-specific patent legislation will encourage rent seeking. Patent law has some balance today in part because different industries have different interests, making it difficult for one interest group to push through changes to the statute. Industry-specific legislation is much more vulnerable to industry capture. It is no accident that the industry-specific portions of the patent law are among the most complex and confusing sections,[191] and that they have had some pernicious consequences.[192] The copyright model -- in which industry-specific rules and exceptions have led to a bloated, impenetrable statute that reads like the tax code[193] -- is hardly one patent law should emulate.[194]

### 2.     General Modifications to Accommodate Policy Concerns

An alternative to industry-specific patent legislation is some sort of general change to the patent law that permits the courts to reach more efficient results in particular industries. General changes are more promising than industry-specific legislation for the reasons we suggest above.[195] In this section, we consider two possible changes to the patent law that might solve the biotechnology-software problem we identify in this article.

---

[190] *Cf.* 35 U.S.C. §103(b), which is irrelevant today largely because general patent standards today reach the same result. *See* In re Ochiai, 71 F.3d 1565 (Fed. Cir. 1995).

[191] In particular, sections 103(b) (biotechnological proceses), 155A (private patent relief), 156 (Hatch-Waxman pharmaceutical patent term extension), and 287 (medical process patents).

[192] The Hatch-Waxman provisions in particular have been used on numerous occasions to violate the antitrust laws. Pharmaceutical patent owners have colluded with putative generic entrants to prevent that company or any other from entering the market. *See, e.g.,* Andrx Pharmaceuticals v. Biovail Corp., __ F.3d __, 2001 WL 855472 (D.C. Cir. July 31, 2001); In re Cardizem CD Antitrust Litigation, 105 F. Supp. 2d 682 (E.D. Mich. 2000).

[193] *See generally* 17 U.S.C. On the unnecessary complexity of the copyright laws, see Jessica Litman, *Revising Copyright Law for the Information Age*, 75 **Or. L. Rev.** 19 (1996); Jessica Litman, *The Exclusive Right to Read*, 13 **Cardozo Arts & Ent. L.J.** 29 (1994).

[194] Indeed, scholars have suggested the opposite -- that copyright law should learn from patent. *See* Mark A. Lemley, *The Economics of Improvement in Intellectual Property Law*, 75 **Tex. L. Rev.** 989 (1997); John Shepard Wiley Jr., *Copyright at the School of Patent*, 58 **U. Chi. L. Rev.** 119 (1991).

### a. Decoupling the PHOSITA

First, courts might decouple the PHOSITA standards for obviousness and enablement, thus allowing the two requirements for patentability to be independently adapted to the incentive requirements of various technologies. The seeds of such an approach may already be latent in established legal doctrines of obviousness and enablement. Recall that the characteristics of the obviousness PHOSITA and those of the enablement PHOSITA are not entirely coterminous; they are measured at different times. Because the level of knowledge for the enablement PHOSITA is measured at the time a patent is filed, rather than as of the date of invention, a larger pool of prior art will frequently be imputed to the knowledge of the enablement PHOSITA. The accumulation of prior art between invention and filing could in theory allow an invention to enjoy both a low threshold of obviousness and a low threshold of disclosure.

It may be that this differential knowledge should be emphasized, in order to decouple the tight reciprocity of obviousness and enablement. In the case of biotechnology, this approach might bring policy expectations into line with doctrinal results. The characteristics of the industry would seem to demand fewer and broader

---

[195] Of course, general modifications may have disproportionate impact on specific industries. Thus, if Wells is right that Internet business methods have low costs of development and are easy to bring to

patents that are easier to obtain, which is to say, patents with a relatively low obviousness threshold, but also a relatively low enablement threshold. These might be provided by the temporal disparity in the two PHOSITAs. Since the body of prior art grows during the period between invention and the filing of a patent, the corpus of knowledge imputed to the enablement PHOSITA will be larger than that imputed to the obviousness PHOSITA. Thus, the level of disclosure required to enable one of ordinary skill at the time of filing could well be lower than that required to enable one of ordinary skill at the time the invention was made, because the enablement PHOSITA is expected to know more. Conversely, an invention may well be non-obvious at the time it is made, although it would not be at the time a patent is filed. This will depend to some extent upon when the invention is considered to be "made" for non-obviousness purposes. If the inventor must rely upon her filing date as the date the invention is made, then the knowledge imputed to the enablement and obviousness PHOSITAs may be coterminous; however, if the date of invention can be related back to an early time of conception, the disparity between the two bodies of prior art may be substantial.

This same effect may hold true for the PHOSITA of the written description requirement, albeit to a lesser extent. The written description requirement substantially overlaps with the degree of disclosure required for enablement, but is likely to require something more. Because the unique purpose of the written description requirement is to demonstrate what the inventor had in his possession at the time the patent was filed, courts have been understandably reluctant to assume that details missing from the disclosure could be supplied by the prior art knowledge imputed to the PHOSITA. Thus,

---

market, *see* Wells, *supra* note __, at 770-72, they are more likely to be found obvious under the standard we propose.

the inventor is less able to rely upon the level of knowledge in the prior art to establish a less stringent requirement for written description. Nonetheless, some courts have suggested that there is some flexibility in the written description requirement, although that has not been the trend in the Federal Circuit biotechnology cases. If the less stringent holdings of those older, cases could be revitalized, the outcome for biotechnology might be brought into line with the policy expectations for that technology.

However, this consonance between policy outcomes and the emphasis on disparities in prior art may be peculiar to biotechnology, and not capable of generalization to other technologies. In particular, reliance upon the differential in prior art between obviousness and enablement may not yield the optimal result in the case of computer software. As we have described, the profile of that industry militates in favor of narrower and more sharply defined patents; or in other words, toward a higher threshold of patentability for both obviousness and disclosure. But, in this case, the differential in prior art between the time of invention and the time of filing pushes in the wrong direction, away from a stringent enablement standard. No matter how high the threshold for obviousness may be set, the passage of time between invention and filing will place more knowledge at the disposal of the PHOSITA at the latter event, favoring less disclosure rather than more. This means that even a decoupled PHOSITA standard won't achieve ideal results in both biotechnology – where the standards may need to be relaxed somewhat – and software – where they need to be tightened.

Consequently, it may be inadequate to rely upon the knowledge differential already found in the PHOSITA standard in order to correct the mismatch of policy

*Is Patent Law Technology-Specific?* Burk & Lemley                                                                      DRAFToutcomes and doctrinal analysis. What may be requited is to decouple the section 103 and section 112 PHOSITA standards altogether, recognizing that the PHOSITA contemplated for purposes of obviousness is simply not the PHOSITA contemplated for purposes of disclosure. Although tight reciprocity of the two standards, mediated by a common PHOSITA construct, makes for an appealing and intellectually elegant doctrinal framework, theoretical esthetics might be required to give way to technological pragmatics. Again, there are precursors latent in the case law that could be developed into such a doctrinal shift; recall that the section 103 PHOSITA has been portrayed by some courts as a bit of an innovator, while the section 112 PHOSITA has not. Certainly the two constructs are conceived as being engaged in very different inquiries, the first seeking some motivation to compile prior art knowledge into an invention, and the second drawing upon prior art knowledge to supplement an invention disclosure. Divorcing the two inquiries could allow each standard the freedom to independently accommodate the incentive needed by a given industry.

### b.     Considering Innovation, Not Just Invention

Second, courts or Congress might modify the standard of obviousness to account for the cost and uncertainty of production and development, not just invention. We think that the divergence between legal and policy outcomes is attributable to the fact that software inventions are much easier to bring to market than biotechnology inventions. The court repeatedly intones the maxim that biotechnology is an "uncertain art."[196] We think, however, that it is not so much invention as product development, production and regulatory approval that is uncertain in biotechnology.

---

[196] *See, e.g.,* In re Vaeck, 947 F.2d 488, 496 (Fed. Cir. 1991) (biotechnology less "predictable" than mechanics or electronics).

From a policy perspective, the result is the same: biotechnological inventions need more incentive than other types of inventions if they are actually to make it to market. This suggests that the lowered obviousness standard found in the Federal Circuit's biotechnology decisions is the correct outcome, but that the court has reached the correct result for the wrong reasons. The justification in biotechnology for a lower standard of obviousness is not, as the Federal Circuit has suggested, the cost or difficulty of invention – numerous commentators have rightly pointed out that the costs of isolating valuable macromolecules are now trivial, and the expectation of success in such an undertaking is high. The justification in biotechnology for a lower obviousness standard is rather that the costs of bringing such products to market are prohibitive. Product development can be encouraged by offering patentees a greater certainty of obtaining the financial rewards from an exclusive patent right.

A number of commentators have suggested that the court should take account of the higher costs and greater uncertainty of innovation in biotechnology by modifying the nonobviousness standard to account for what might be called "economic nonobviousness."[197] Karen Boyd argues that permitting evidence of economic nonobviousness would improve biotechnology policy, and that the court could start with a rough assumption that all biotechnological inventions are high-cost inventions for which a finding of economic nonobviousness would be appropriate.[198] Rob Merges makes the more general argument that the courts should reduce the obviousness standard either in cases in which the invention itself is uncertain, or in which the commercialization process is long and costly.[199]

---

[197] Boyd, *supra* note __, at 311.

[198] *See id*. at 337-339.

We think Merges and Boyd are on the right track in pointing to the cost and uncertainty of innovation as a whole – not just invention – as relevant factors in the nonobviousness determination.[200] But we also think that courts in the biotechnology obviousness cases have effectively done this already, implicitly accommodating the cost of development in cases like *Bell* and *Deuel*.[201] Nonetheless, there are at least three important reasons to consider innovation cost expressly, rather than by modifying existing law *sub silentio*. First, it is rarely wise for courts to conceal the reasons for their decisions. The integrity of the patent law can only be enhanced by a forthright discussion of the values served by the decisions in *Bell* and *Deuel*.

Second, considering innovation cost expressly makes it clear that the level of skill in the art is not what is driving decisions like *Bell* and *Deuel*. As a result, the Federal Circuit should not feel constrained to decide section 112 cases in accordance with the distorted view of biotechnology presented in those cases. In particular, the court's syllogism in *Eli Lilly* – if nonobvious, then not sufficiently described[202] – could safely be rejected if the reason for nonobviousness were not related to the level of skill in the art.

---

[199] Merges, *Uncertainty, supra* note __, at 34-36 (technical uncertainty), 48-49 (cost of development).

[200] *Cf.* A. Samuel Oddi, *Beyond Obviousness: Invention Protection in the Twenty-First Century*, 38 **Am. U. L. Rev.** 1097, 1116 (1989) (arguing that cost should drive the nonobviousness calculation).

[201] *Accord* Boyd, *supra* note __, at 342 ("By any honest interpretation of the nonobviousness statute, the decisions in *Deuel* and *Bell* are wrong"; nonetheless endorsing the results for policy reasons).

[202] *See* Regents of the University of California v. Eli Lilly & Co., 119 F.3d 1559 (Fed. Cir. 1997).

Finally, explicit consideration of development cost and uncertainty can work both ways. Commentators have generally focused on economic nonobviousness as a reason to enhance protection in industries like biotechnology.[203] While most biotechnological inventions would benefit from such consideration because of their high cost and uncertain development process, in the case of software development the opposite is true. Software inventions tend to have a quick, cheap, and fairly straightforward post-invention development cycle. Most of the work in software development occurs in the initial coding, not in development or production. The lead time to market in the software industry tends to be short. The capital investment requirement for software development is relatively low -- mostly hiring personnel, not building laboratories or manufacturing infrastructure. Debugging and test-marketing is tedious and occasionally time-consuming, but does not rival the cost of stringent safety testing and agency oversight necessary in the biotechnology and pharmaceutical industries. As a result, we believe the application of economic principles of nonobviousness should raise the obviousness bar in the software industry, just as it is lowered in the biotechnology industry. More generally, economic nonobviousness is a principle that must be applied across the board, not just to benefit patentees in a particular industry.

Judicial consideration of the costs and uncertainty of the product development process is feasible. Courts already consider a variety of "secondary considerations" of nonobviousness, despite the absence of any statutory authority for doing so. While the Court in *Graham v. John Deere* decided merely that secondary considerations "may have relevancy,"[204] the Federal Circuit regularly speaks of them as a required part of any obviousness inquiry.[205] And the secondary considerations that have been used tend to be economic indicia of a patent's value: the commercial success of a patent, long-felt need for the invention and the failure of others to produced it, market response to the invention by copying or acquiescence, and simultaneous invention.[206] One may reasonably question the economic value of some of this evidence,[207] but it is clear that the courts are already willing to consider economic indicia of nonobviousness of their own volition in a variety of patent cases. As a result, introducing evidence of development cost and uncertainty will not substantially disrupt the existing processes for determining patent validity.

---

[203] *See* Merges, *Uncertainty, supra* note __; Boyd, *supra* note __.

A case-specific focus on development cost as a plus-factor also avoids many of the problems with industry-specific patent legislation discussed above.[208] Because courts could easily adopt this test themselves, no legislation is necessary. This not only avoids administrative costs, but it also avoids the very real problem of industry rent-seeking. Further, because it is a general policy, it can be applied to all types of patent cases. While the results will be tailored to the characteristics of particular industries, the legal rules will not be. Indeed, by creating a framework in which courts can take account of legitimate policy concerns, the Federal Circuit may actually be able to reduce the judicial "drift" towards industry-specific legal rules that we observed in Part I.

---

[204] 383 U.S. 1, 18 (1966).

[205] *See, e.g.,* Greenwood v. Haitori Seiko Co., 900 F.2d 238, 241 (Fed. Cir. 1990).

[206] *See id*. considering the first four factors listed; *cf.* Hybritech v. Monoclonal Antibodies, Inc., 802 F.2d 1367, n.2 (Fed. Cir. 1986) (simultaneous invention may be a relevant factor, but wasn't in that case).

[207] *See, e.g.,* Robert P. Merges, *Economic Perspectives on Innovation: Patent Standards and Commercial Success*, 76 **Calif. L. Rev.** 803 (1988); Edmund Kitch, *Graham v. John Deere Co.: New Standards for Patents*, 1966 **Sup. Ct. Rev.** 293; Rochelle Cooper Dreyfuss, *The Federal Circuit: A Case Study in Specialized Courts*, 64 **N.Y.U. L. Rev.** 1 (1989) (all criticizing reliance on commercial success).

[208] *See supra* notes __-__ and accompanying text.

## Conclusion

Patent law is becoming technology-specific. The legal rules applied to biotechnology cases bear less and less resemblance to those applied in software cases. While there are good policy reasons to treat the two industries differently, the current legal rules are not expressly informed by the economics of the industries, but by an ad hoc combination of judicial policymaking and stare decisis. Not surprisingly, they don't reflect optimal patent policy in either biotechnology or software. We have offered some explanations for this phenomenon, along with two specific suggestions – decoupling the obviousness and enablement standards and adding development cost to the secondary considerations of nonobviousness. These suggestions will help optimize patent policy in general and biotechnology and software law in particular.